\documentclass[aps,twocolumn,prb,superscriptaddress]{revtex4-1}
\usepackage{times}
\usepackage{amsmath}
\usepackage{graphicx}
\usepackage[colorlinks=true,citecolor=blue,urlcolor=blue,anchorcolor=blue,linkcolor=blue]{hyperref}

\begin{document}

\title{Prediction of superconductivity and topological aspects in single-layer $\beta$-Bi$_{2}$Pd}

\author{Peng-Fei Liu}
\thanks{These two authors contributed equally to this work.}
\affiliation{Spallation Neutron Source Science Center, Institute of High Energy Physics, Chinese Academy of Sciences, Dongguan 523803, China}

\author{Jingyu Li}
\thanks{These two authors contributed equally to this work.}
\affiliation{Institute for Computational Materials Science, School of Physics and Electronics, International Joint Research Laboratory of New Energy Materials and Devices of Henan Province, Henan University, Kaifeng 475004, China}

\author{Xin-Hai Tu}
\affiliation{Spallation Neutron Source Science Center, Institute of High Energy Physics, Chinese Academy of Sciences, Dongguan 523803, China}

\author{Huabing Yin}
\affiliation{Institute for Computational Materials Science, School of Physics and Electronics, International Joint Research Laboratory of New Energy Materials and Devices of Henan Province, Henan University, Kaifeng 475004, China}

\author{Baisheng Sa}
\affiliation{Key Laboratory of Eco-materials Advanced Technology, College of Materials Science and Engineering, Fuzhou University, Fuzhou 350108, China}

\author{Junrong Zhang}
\affiliation{Spallation Neutron Source Science Center, Institute of High Energy Physics, Chinese Academy of Sciences, Dongguan 523803, China}

\author{David J. Singh}
\email{singhdj@missouri.edu}
\affiliation{Department of Physics and Astronomy, University of Missouri, Columbia, Missouri 65211, USA}

\author{Bao-Tian Wang}
\email{wangbt@ihep.ac.cn}
\affiliation{Spallation Neutron Source Science Center, Institute of High Energy Physics, Chinese Academy of Sciences, Dongguan 523803, China}
\affiliation{Collaborative Innovation Center of Extreme Optics, Shanxi University, Taiyuan, Shanxi 030006, China}

\date{\today}

\begin{abstract}
Topological superconductors,
characterized by topologically nontrivial states residing in a superconducting gap,
are a recently discovered class of materials having Majorana Fermions.
The interplay of superconductivity and topological states give rise to opportunities
for achieving such topological superconductors in condensed matter systems.
Up to now, several single-material topological superconductors in this form
have been theoretically predicted and experimentally confirmed.
Here, using the first-principles calculations, we study the superconducting single-layer $\beta$-Bi$_{2}$Pd.
The electronic density of states near Fermi level of this monolayer
are dominated by the Bi-p and Pd-d orbitals, forming a two-band Fermi surface with multi-class sheets.
The presence of soft phonon bands, in cooperation with the electron susceptibility, account for
electron-phonon superconductivity of single-layer $\beta$-Bi$_{2}$Pd.
With the centrosymmetric structure, single-layer $\beta$-Bi$_{2}$Pd
possesses a continuous gap over the whole Brillouin zone and topological Dirac-like states at its one-dimensional boundary.
The present findings would lead to the expectation of
one-dimensional topological superconductivity and Majorana bound states in monolayer candidate of $\beta$-Bi$_{2}$Pd with intrinsic full-gap superconductivity.
\end{abstract}

\maketitle
\section{Introduction}

Two dimensional (2D) superconductors, while known, as in Pb thin films
\cite{Zhang2010,Eom2006},
have attracted much recent interest due to discoveries such as evidence for very
high critical temperatures in oxide supported monolayer FeSe \cite{He2013,Ge2014},
the superconductivity of bilayer twisted graphene \cite{cao2018},
the finding of unconventional behavior in strong spin orbit split cases
\cite{nam}
and the finding of undiminished high temperature superconductivity in cuprate monolayers
\cite{yu2019}.
Furthermore, 2D materials enable unique device geometries due to the ability to fabricate
stacked structures and manipulate properties.
A particular interest is in devices related to quantum information.
In this regard,
superconductors with nontrivial topological states are of interest
due to the possibility of achieving Majorana Fermions \cite{Majorana1937,nayak}.

There have been a number of reports observing
superconductivity in topologically nontrivial systems
with signatures of unusual superconductivity.
These include
Cu- or Sr-doped Bi$_{2}$Se$_{3}$\cite{Hor2010,Matano2016,Liu2015},
Bi$_{2}$Se$_{3}$/NbSe$_{2}$ heterostructures\cite{Xu2015,Sun2016},
Bi$_{2}$Se$_{3}$ or Bi$_{2}$Te$_{3}$ on Bi$_{2}$Sr$_{2}$CaCu$_{2}$O$_{8+\delta}$\cite{Zareapour2012},
atomic Fe chains on Pb(110)\cite{NadjPerge2014},
and Nb structures on (Cr$_{0.12}$Bi$_{0.26}$Sb$_{0.62}$)$_{2}$Te$_{3}$ thin film\cite{He2017}.
These generally involve production of superconductivity by
doping topological materials or by proximity effects.
While these are effective strategies
\cite{PhysRevLett.111.087001,Zhao2015},
it is desirable to identify intrinsic topological superconductors, particularly
with chemical stoichiometry
\cite{PhysRevB.39.12743,PhysRevB.64.064424,Sakano2015,Iwaya2017,Guan2016,Wang2020},
as well as 2D versions.

Evidence for Majorana zero modes was found in some intrinsic superconductors such as
2M phase WS$_{2}$\cite{Yuan2019,fang2019discovery},
$T_{d}$-MoTe$_{2}$\cite{guguchia2017signatures}, BiPd\cite{Sun2015},
$\beta$-Bi$_{2}$Pd\cite{Sakano2015},
and PbTaSe$_{2}$\cite{Chang2016}.
Of these $\beta$-Bi$_2$Pd is of particular interest due to its naturally layered crystal structure [see Fig. \ref{fgr:fig-1}(a)],
which gives rise to the possible 1D topological superconductivity in low-dimensional $\beta$-Bi$_{2}$Pd.
Layered $\beta$-Bi$_{2}$Pd is centrosymmetric (space group I4/mmm) with superconducting critical temperature, $T_c$=5.4 K
\cite{PhysRevB.93.144502,Iwaya2017,Imai2012}.
Within the fully anisotropic Migdal-Eliashberg formalism\cite{PhysRevB.87.024505},
$\beta$-Bi$_2$Pd displays
phonon-mediated superconductivity with a single anisotropic superconducting gap\cite{PhysRevB.95.014512}, and
the quantum electronic stress induced by quantum confinement can effectively enhance its $T_c$ \cite{PhysRevB.100.104527}.
Furthermore, this compound has nontrivial band topology with topologically protected surface states
and Rashba-like spin splitting\cite{Sakano2015,Wang2017}.
This makes it a promising candidate for investigation of topological superconductivity,
independent of doping or proximity effects\cite{Li2019}.
For $\beta$-Bi$_{2}$Pd, its layered crystal structure is composed of Bi$_2$Pd sheets, consisting of Pd atoms between Bi planes
\cite{Wang2017,Xu2019,GUAN20191215,PhysRevB.97.134505}.
Since Bi-Bi bonding between the Bi$_2$Pd sheets is weak\cite{Wang2017},
the material is amenable to thin film growth and can form monolayers\cite{Denisov2017}.
In the meantime, theoretical calculations verify that films of Bi$_2$Pd could harbor topological surface states with Dirac- and
Rashba-like band dispersions depending on the thickness\cite{Wang2017}, while
experiments observe that a few films are superconducting with evidence for the topological behavior\cite{GUAN20191215}.
These demonstrate that $\beta$-Bi$_{2}$Pd
could provide a reliable platform for achieving the long-sought topological superconductor in the low-dimensional limit.

In the previous studies, much attention has been paid to the electronic bands, phonons, superconductivity, and topology in  bulk and thin films of $\beta$-Bi$_{2}$Pd
\cite{Sakano2015,PhysRevB.95.014512,PhysRevB.100.104527,Wang2017,Xu2019,GUAN20191215,PhysRevB.97.134505}.
The data on single-layer $\beta$-Bi$_{2}$Pd is still scarce in the literature.
Here we use first-principles calculations to show that single-layer $\beta$-Bi$_2$Pd is a 2D electron phonon
superconductor with topological edge states at its 1D boundary.
Specifically, we start with the layered structure and present a theoretical study
of the electronic structure and superconductivity of single-layer
$\beta$-Bi$_{2}$Pd via state-of-the-art first-principles calculations.
We find that single-layer $\beta$-Bi$_{2}$Pd has a two-band Fermi surface with mixed Bi-p/Pd-d orbital character.
Detailed calculations of the electron phonon Eliashberg spectral function, $\alpha^2F(\omega)$ show
that electron phonon superconductivity is present.
Strong contributions
are found both for optic phonons that intersect the acoustic branches with strong Bi contributions,
as well as higher frequency modes with Pd character.
Taking together the superconductivity and its intrinsic edge states of single-layer $\beta$-Bi$_{2}$Pd,
our work points out a real material that
is a promising platform for studying 1D topological superconductivity and Majorana physics.

\section{Computational methods}

Our calculations were performed within density functional theory
with norm-conserving pseudopotentials\cite{PhysRevB.43.1993,FUCHS199967},
as implemented in the Quantum-ESPRESSO package\cite{qe2009,qe2017}.
We used the Perdew-Burke-Ernzerhof\cite{PhysRevLett.77.3865} generalized gradient approximation.
All self-consistent calculations were performed with a planewave kinetic energy cutoff of 80 Ry.
We used a Brillouin zone (BZ) sampling based on a 32$\times$32$\times$1 $\textbf{\emph{k}}$-mesh.
A Methfessel-Paxton smearing of 0.02 Ry was employed for these calculations.
The internal atomic positions were fully relaxed
with a threshold of 10 meV/\AA{} for the forces.
We did calculations using a supercell, of length
15 \AA{} along the \emph{z} direction.

The phonon spectrum and electron-phonon coupling (EPC) strength $\lambda$ were calculated
with density functional perturbation theory\cite{baroni2001phonons}
on a 8$\times$8$\times$1 $\textbf{\emph{q}}$-mesh.
Here, the mode-resolved magnitude of the EPC $\lambda_{\textbf{\emph{q}}\nu}$
is calculated according to the Migdal-Eliashberg theory\cite{grimvall1981electron,giustino2017electron}
by
\begin{align}
\lambda_{\textbf{\emph{q}}\nu}=\frac{\gamma_{\textbf{\emph{q}}\nu}}{\pi\!hN(E_{\mathrm{{F}}})\omega_{\textbf{\emph{q}}\nu}^{2}},
\end{align}
where $\gamma_{\textbf{\emph{q}}\nu}$ is the phonon linewidth,
$\omega_{\textbf{\emph{q}}\nu}$ is the phonon frequency,
and $N$(\emph{E}$_{\mathrm{{F}}}$) is the electronic density of states at the Fermi level.
The $\gamma_{\textbf{\emph{q}}\nu}$ can be obtained with
\begin{align}
\gamma_{\textbf{\emph{q}}\nu}=\frac{2\pi\omega_{\textbf{\emph{q}}\nu}}{\Omega_{\rm{BZ}}}\sum_{\textbf{\emph{k}},n,m}|\rm{g}_{\textbf{\emph{k}}n,\textbf{\emph{k}}+\textbf{\emph{q}}m}^{\nu}|^{2}\delta(\varepsilon_{\textbf{\emph{k}}n}-\varepsilon_{F})\delta(\varepsilon_{\textbf{\emph{k}}+\textbf{\emph{q}}m}-\varepsilon_{F}),
\end{align}
where $\Omega_{\rm{BZ}}$ is the volume of BZ,
$\varepsilon_{\textbf{\emph{k}}n}$ and $\varepsilon_{\textbf{\emph{k}}+\textbf{\emph{q}}m}$
denote the Kohn-Sham energy, and $\rm{g}_{\textbf{\emph{k}}n,\textbf{\emph{k}}+\textbf{\emph{q}}m}^{\nu}$
is the EPC matrix element.
The $\rm{g}_{\textbf{\emph{k}}n,\textbf{\emph{k}}+\textbf{\emph{q}}m}^{\nu}$,
which can be determined self-consistently by the linear response theory,
describe the probability amplitude for the scattering
of an electron with a transfer of crystal momentum $\textbf{\emph{q}}$\cite{allen1975transition}.
The Eliashberg electron-phonon spectral function $\alpha^{2}F(\omega)$ is then
\begin{align}\label{equ:a2f}
\alpha^{2}F(\omega)=\frac{1}{2\pi\!N(E_{\mathrm{{F}}})}\sum_{\textbf{\emph{q}}\nu}\frac{\gamma_{\textbf{\emph{q}}\nu}}{\omega_{\textbf{\emph{q}}\nu}}\delta(\omega-\omega_{\textbf{\emph{q}}\nu}).
\end{align}
The EPC constant $\lambda$ can be calculated either by the summation of the EPC constant $\lambda_{\textbf{\emph{q}}\nu}$ in the full BZ for all phonon modes or by the integral of the Eliashberg spectral function $\alpha^{2}F(\omega)$\cite{osti_7354388} as
\begin{align}\label{equ:lambda}
\lambda(\omega) = \sum_{\textbf{\emph{q}}\nu} \lambda_{\textbf{\emph{q}}\nu} = 2\int_{0}^{\omega}\frac{\alpha^{2}\emph{F}(\omega)}{\omega}d\omega.
\end{align}

We use the McMillan-Allen-Dynes formula to obtain the superconducting transition temperature, $T_c$,
from the calculated EPC constant $\lambda$,
\begin{align}\label{equ:tc}
T_{c}=f_1f_2\frac{\omega\mathrm{{_{log}}}}{1.2}\mathrm{{exp}\left[-\frac {1.04(1+\lambda)}{\lambda-\mu^{*}(1+0.62\lambda)}\right]},
\end{align}
where $\mu^{*}$ is the effective screened Coulomb repulsion constant, $\omega_{log}$ is the logarithmic average frequency,
\begin{align}\label{equ:logw}
\omega\rm{_{log}}=exp\left[\frac{2}{\lambda}\int_{0}^{\infty}
\frac{\emph{d}\omega}{\omega}\alpha^{2}\emph{F}(\omega)\rm{log}\omega\right],
\end{align}
and
$f_{i}$ is the correction factor when $\lambda>$ 1.3\cite{PhysRevB.12.905}.
In our calculation, $\mu^{*}$ = 0.1 and $f_1f_2$ = 1 were used.
This is based on parameters used in bulk $\beta$-Bi$_{2}$Pd previously\cite{PhysRevB.95.014512}.

In order to analyze the  electronic susceptibility and edge states of single-layer $\beta$-Bi$_{2}$Pd,
we used a tight binding Hamiltonian, constructed from the first-principles Bloch functions.
This was done by projecting
Bloch states onto maximally localized Wannier functions (MLWFs)\cite{RevModPhys.84.1419}
with the Wannier90 package\cite{MOSTOFI20142309,Pizzi_2020}.
In the model, the MLWFs are derived from atomic Bi-s, Bi-p, and Pd-d orbitals.
Using the MLWFs, the edge states are calculated from the imaginary part of the surface Green's function
\cite{Sancho1985}
as provided by the WannierTools package\cite{Wu2018}.

\section{Results and Discussion}

\subsection{Crystal and band structures}

\begin{figure}[htbp]
	\centering
	\includegraphics[width=0.8\linewidth]{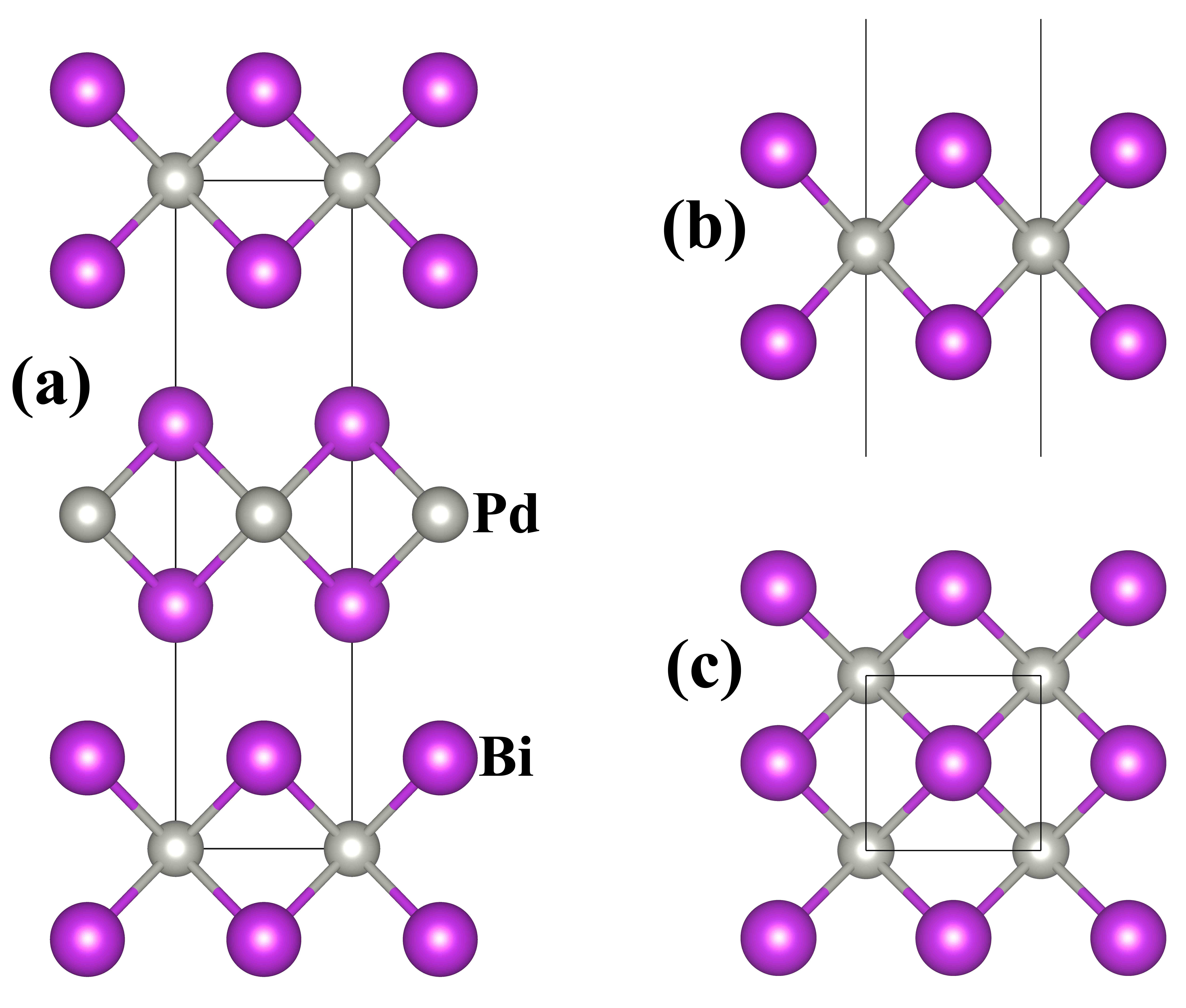}
	\caption{(a) Side view of bulk $\beta$-Bi$_{2}$Pd crystal. (b) Side and (c) top views of single-layer $\beta$-Bi$_{2}$Pd. The purple and gray balls indicate Bi and Pd atoms, respectively. Unit cells are indicated by the solid black line.}
	\label{fgr:fig-1}
\end{figure}
\begin{figure}[htbp]
	\centering
	\includegraphics[width=0.8\linewidth]{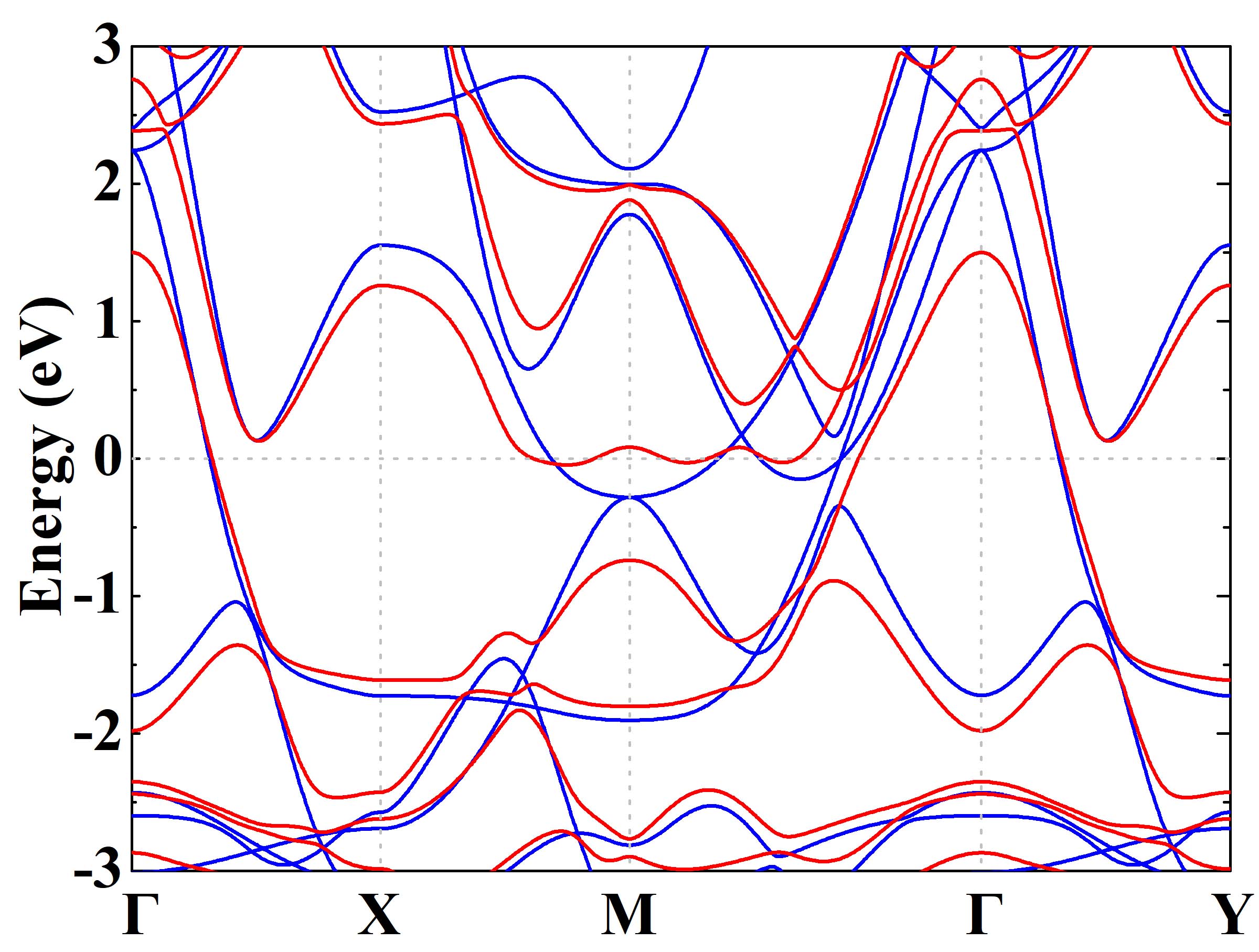}
	\caption{The band structures of single-layer $\beta$-Bi$_{2}$Pd with (blue) and without (red) SOC.
	The high-symmetry points, $\Gamma$, X, M, and Y are (0, 0), (1/2, 0), (1/2, 1/2), and (0, 1/2), respectively.
	The Fermi level is taken as the energy zero.}
	\label{fgr:fig-2}
\end{figure}

\begin{figure}[htbp]
	\centering
	\includegraphics[width=1\linewidth]{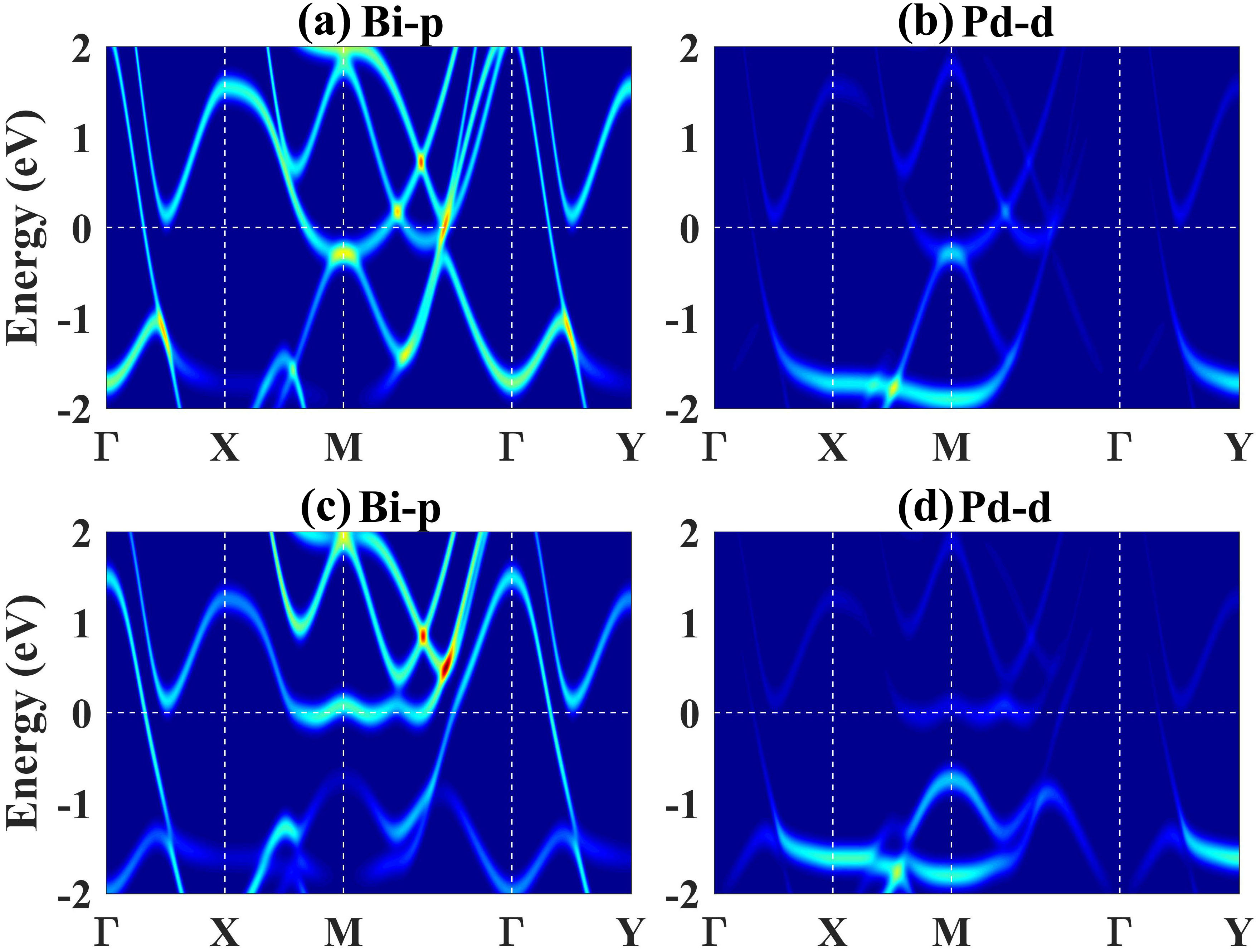}
	\caption{The orbital projected band structure of single-layer $\beta$-Bi$_{2}$Pd weighted by (a) Bi-p, and (b) Pd-d orbitals. Panels (c) and (d) are the same as panels (a) and (c), respectively, but with SOC.	
	The Fermi level is taken as the energy zero.}
	\label{fgr:fig-3}
\end{figure}

As mentioned, $\beta$-Bi$_{2}$Pd is a layered compound with tetragonal centrosymmetric
space group, \emph{I}4/\emph{mmm},
with Bi on site 4e and Pd on 2a, as shown in Fig. \ref{fgr:fig-1}(a)
\cite{Wang2017,Xu2019,GUAN20191215,PhysRevB.97.134505}.
The structure can be described as a close packed stacking of triple layers each consisting of
a square Pd atomic layer sandwiched between two square Bi atomic layers to
yield layering sequence, Bi-Pd-Bi.
As discussed in prior work, the bonding within a triple layer, particularly the Pd-Bi
bonding is substantially stronger than the Bi-Bi bonding between different triple layer units
\cite{Wang2017}.
Thus monolayers consisting of triple layer sheets can be made using molecular beam epitaxy\cite{Denisov2017},
similar to many 2D metal chalcogenides and related materials\cite{zhou2018library}.

We construct the monolayer compound starting with the bulk structure.
The primitive cell of monolayer
$\beta$-Bi$_{2}$Pd is square with two Bi and one Pd atoms in one unit.
The optimized lattice constant is calculated to be 3.352 \AA,
which is slightly smaller than the experimental lattice constant (3.362 \AA)
of bulk $\beta$-Bi$_{2}$Pd\cite{PhysRevB.95.014512}.
The Bi-Pd bond length is stretched to 2.999 \AA{} compared with measured bulk value.
This leads to a monolayer thickness measured from Bi in the top layer to Bi in the bottom layer of 3.674 \AA.

The electronic structure near the Fermi level is metallic in Fig. \ref{fgr:fig-2}. Without spin orbit coupling (SOC),
it clearly shows the valence band and conduction
band touch at the M point and along the $\Gamma$-M line.
Two bands cross the Fermi level leading to the metallic nature.
SOC leads to splitting of degenerate bands around the Fermi level and gives a large gap of 0.77 eV at M point.
Importantly, the bands crossing the Fermi level are gapped from each other, so that
there is a continuous gap between them over the whole BZ (Figs. \ref{fgr:fig-2} and S1), in agreement with the previous results\cite{Zhu2020PdBi}.
We also have checked the band structures with HSE06 hybrid functional\cite{Heyd2003} in Fig. S3.
Figures \ref{fgr:fig-3}(a-b) depict the orbital projected band structure of single-layer $\beta$-Bi$_{2}$Pd
along the main high-symmetry directions of the BZ without the inclusion of SOC.
The Fermi level is derived from hybridized bands of primarily
Bi p and Pd d character. In the present case, the gapping is large, reflecting the strong effect of SOC on Bi p orbitals.

\subsection{Phonons and electron phonon superconductivity}
\begin{figure}[htbp]
	\centering
	\includegraphics[width=1\linewidth]{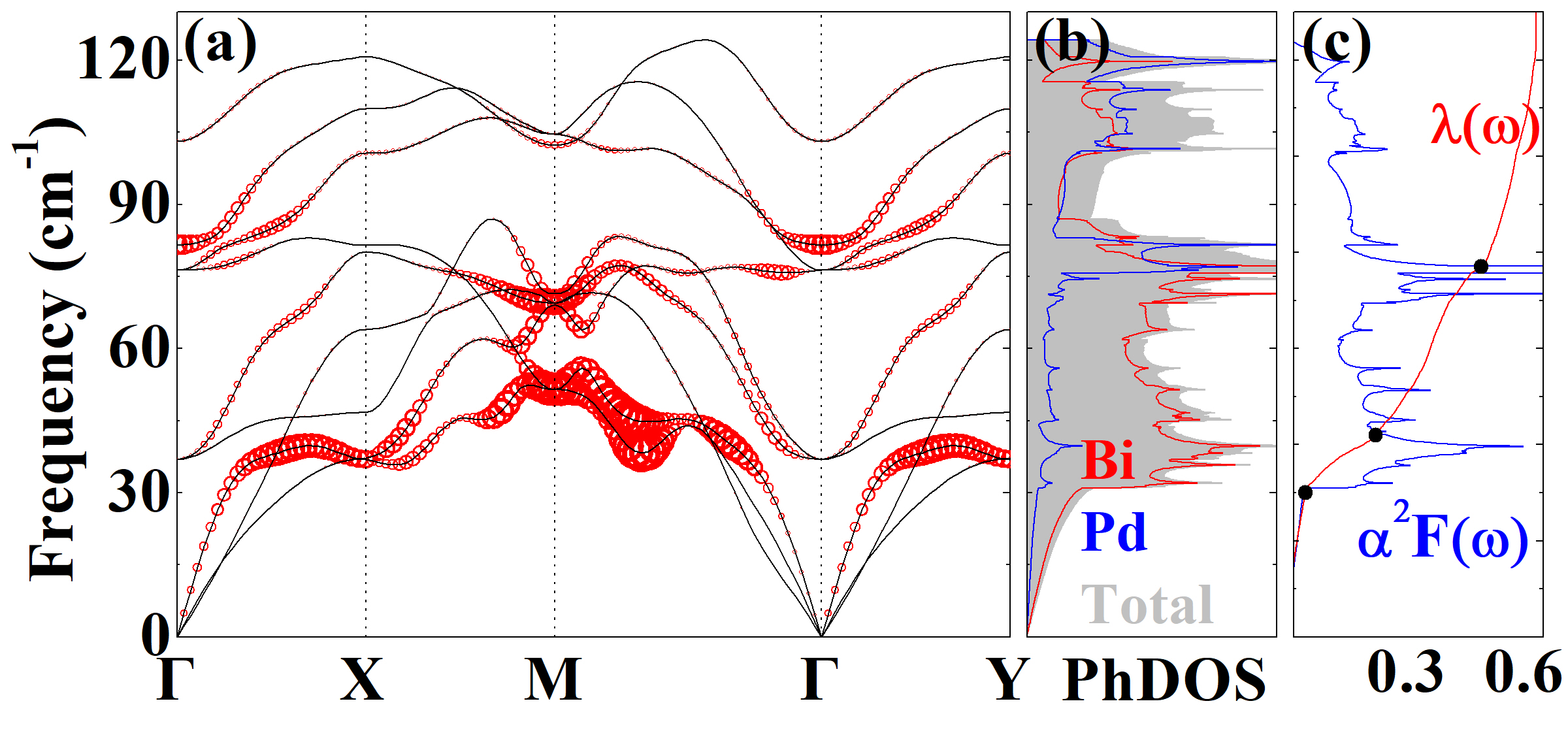}
	\caption{(a) Calculated phonon dispersion along high symmetry lines for monolayer $\beta$-Bi$_{2}$Pd. The size of red circles is proportional to the magnitude of $\lambda_{qn}$. (b) Projected phonon density of states for single-layer $\beta$-Bi$_{2}$Pd. (c) Frequency-dependent Eliashberg spectral function $\alpha^2F(\omega)$ and cumulative frequency-dependent EPC function $\lambda(\omega)$ for single-layer $\beta$-Bi$_{2}$Pd. The black dots in the panel (c) are at 30, 42, and 77 cm$^{-1}$, respectively.}
	\label{fgr:fig-4}
\end{figure}

\begin{table}[htbp]
	\small
	\caption{\ $N$(\emph{E}$_{\mathrm{{F}}}$) (in unit of states per spin per eV per cell), $\omega _{log}$ (in K), $\lambda$, and $T_{c}$ (in K) for $\beta$-Bi$_{2}$Pd. The values in parentheses are calculated with SOC.}
	\label{tbl:tbl-2}
    \begin{center}
  	\begin{tabular}{ccccccccccccccccccccccccccccccccccccccccc}
  		\hline
		\hline
		Comp. & $N$(\emph{E}$_{\mathrm{{F}}}$) & $\omega _{log}$ & $\lambda$ & $T_{c}$ & Ref. \\
		\hline
		2D  & 0.66 & 73.93 & 0.63 & 1.95  & Ours \\
		3D  & 0.66 (0.79) & 101.72 (68.53) & 0.77 (0.97) & 4.40 (4.55) & \cite{PhysRevB.95.014512} \\
        3D  & 1.331 & 83.89 & 0.81 & 4.04 & \cite{PhysRevB.100.104527} \\
		\hline
	\end{tabular}
    \end{center}
\end{table}

The phonon dispersion, presented in Fig. \ref{fgr:fig-4}(a), has no unstable modes.
This confirms the dynamic stability of single-layer $\beta$-Bi$_{2}$Pd.
There are two low-energy optical branches that start just above 30 cm$^{-1}$ at $\Gamma$
and intersect the acoustic branches going towards the zone boundary.
These show substantial electron phonon coupling in the zone, and especially one may note an apparent
Kohn anomaly along $\Gamma$-M near the M point, as well as substantial coupling in the next higher set
of modes at M where the extrapolated acoustic modes reach the zone boundary.
There is also substantial coupling evident in the second set of optic modes starting near 80 cm$^{-1}$ near $\Gamma$.
The phonon density of states (PhDOS) is shown in Fig. \ref{fgr:fig-4}(b).
Both types of atoms are involved in the entire frequency range.
However, the lower frequency phonons mainly stem from the vibration of Bi atoms,
while the higher optical modes above 75 cm$^{-1}$ are from the hybridized vibrations of Pd and Bi atoms.
This is consistent with the behavior of bulk $\beta$-Bi$_{2}$Pd\cite{PhysRevB.95.014512}
and is consistent also with expectations from the heavier mass of Bi relative to Pd as well as the relatively
strong bonding of the Pd, found in prior work.

We now turn to superconductivity, based on the calculated
Eliashberg spectral function $\alpha^2F(\omega)$ and the cumulative EPC strength $\lambda(\omega)$.
As shown in Fig. \ref{fgr:fig-4}(c),
four regions can be distinguished in $\alpha^2F(\omega)$ as indicated by black dots.
These are
a low-energy region that extends up to 30 cm$^{-1}$ (energy region I),
two intermediate regions from 30 to 42 cm$^{-1}$ (energy region II) and from 42 to 77 cm$^{-1}$ (III),
and a high-energy region above 77 cm$^{-1}$ (IV).
Although there are contributions to the total EPC from all phonon modes over the entire range,
the relative contributions of each region are considerably different.
The acoustic phonons of region I account for only approximately 6\% of the total EPC strength of $\lambda$ = 0.63.
As mentioned, phonons in region II
include a Kohn anomaly along the $\Gamma$-M line [Fig. \ref{fgr:fig-4}(a)] with substantial electron phonon coupling.
This region provides an approximately 29\% contribution to the total $\lambda$.
Region III makes the main contribution to the EPC, accounting for approximately 43\% of the total $\lambda$.
The higher frequency phonons (IV) have a smaller, but non-negligible contribution of 22\%.
Unlike several other 2D materials,
such as borophene\cite{Zhao2018}, Cu$_{2}$Si\cite{Yan2019}, and Mo$_{2}$C\cite{Zhang2017},
the EPC induced by high-frequency phonons is important.
As shown in Fig. \ref{fgr:fig-4}(b), the phonons of regions II and III are
mainly from the vibrations of Bi atoms and
dominate the EPC of single-layer $\beta$-Bi$_{2}$Pd.

This provides the ingredients for calculating the transition temperature $T_c$.
With a previously used value $\mu^{*}$=0.1, we obtain a superconducting transition temperature
$T_{c}$=1.95 K at ambient pressure, which is smaller than that (4.40 K) of
bulk $\beta$-Bi$_{2}$Pd\cite{PhysRevB.95.014512,PhysRevB.100.104527} as obtained by
similar methods (see Table \ref{tbl:tbl-2}).
The bulk value is in accord with experiment.
Using a higher Coulomb repulsion parameter, $\mu^*$=0.13, lowers $T_c$ as expected, but a sizable
value of 1.4 K is still obtained. Thus it is highly likely that monolayer $\beta$-Bi$_2$Pd is
a superconductor with $T_c$ in the range of 1 K or above.

\subsection{Fermi surface and electronic susceptibility}

\begin{figure}[htbp]
	\centering
	\includegraphics[width=0.8\linewidth]{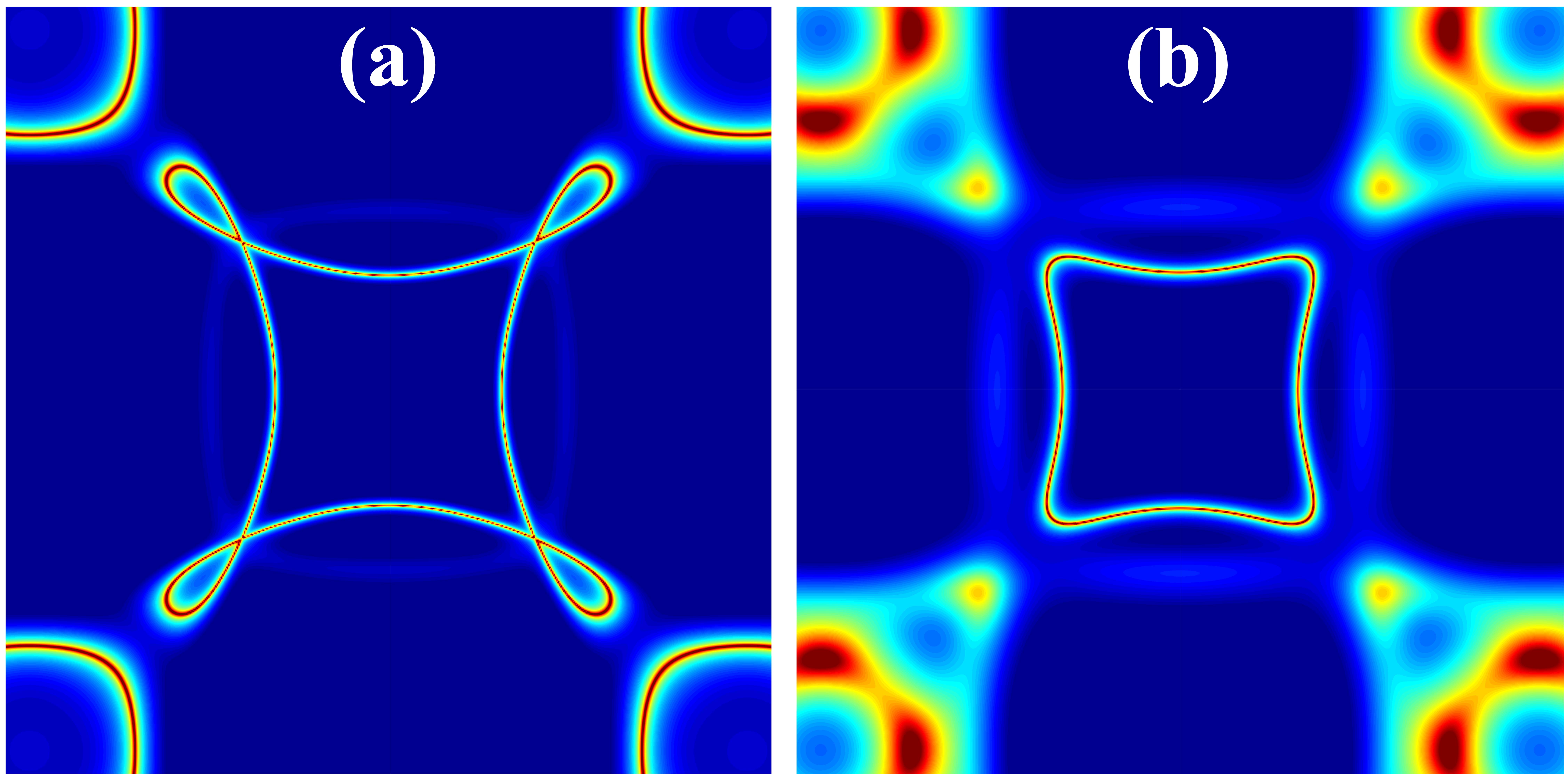}
	\caption{Fermi surfaces (a) without and (b) with SOC formed by two bands across the Fermi level in monolayer
	$\beta$-Bi$_{2}$Pd.}
	\label{fgr:fig-5}
\end{figure}

\begin{figure}[htbp]
	\centering
	\includegraphics[width=0.8\linewidth]{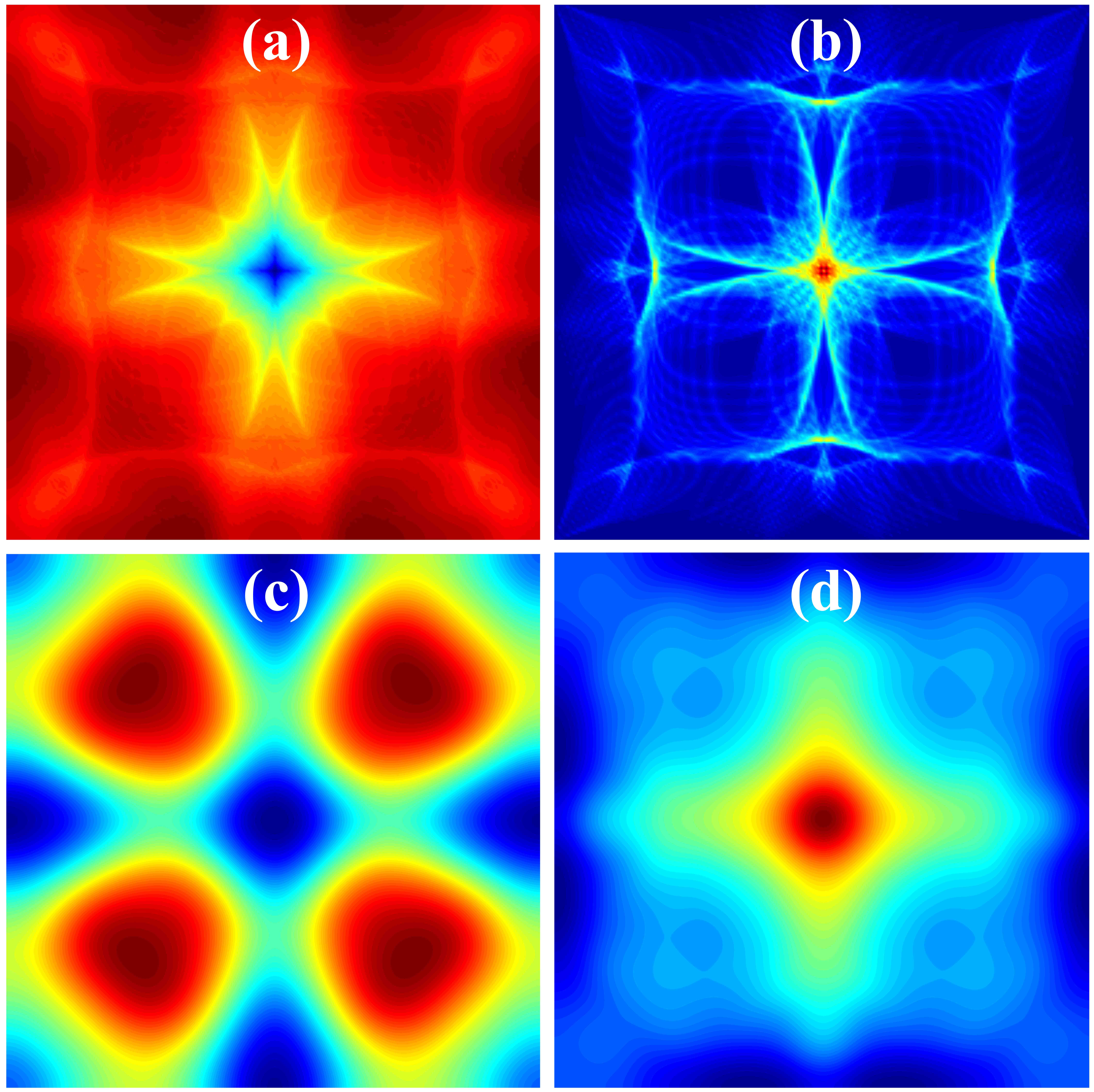}
	\caption{ (a) Real part (static or bare) $\chi'(\textbf{q})$
	and (b) imaginary part (Fermi nesting function) $\chi''(\textbf{q})$ of the electronic susceptibility for single-layer $\beta$-Bi$_{2}$Pd. (c) and (d) are the corresponding results calculated with SOC. Different colors represent the strength of $\chi'(\textbf{q})$ and $\chi''(\textbf{q})$.}
	\label{fgr:fig-6}
\end{figure}

\begin{figure*}[htbp]
	\centering
	\includegraphics[width=1\linewidth]{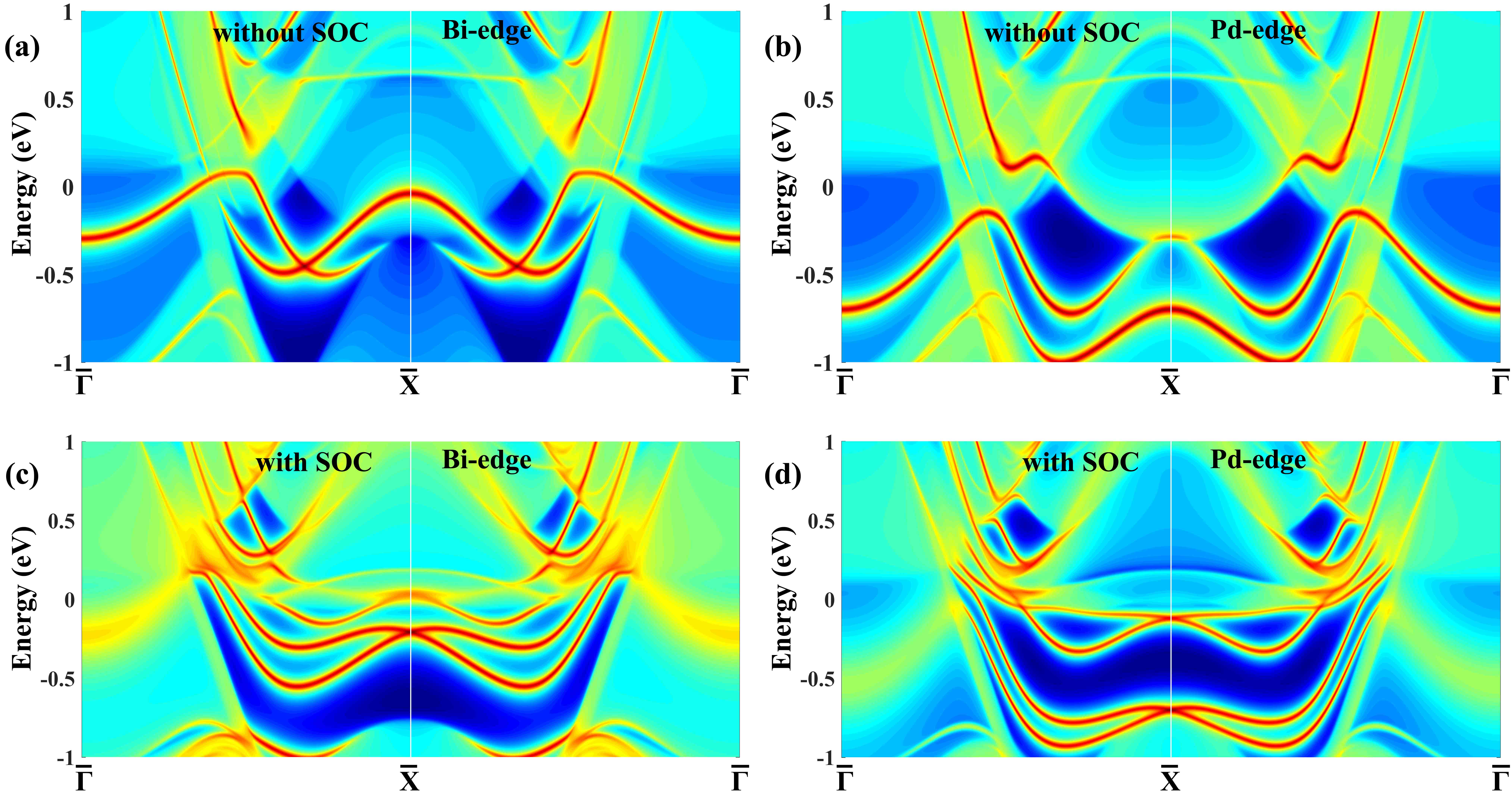}
	\caption{Calculated topological edge states of single-layer $\beta$-Bi$_{2}$Pd for (a) Bi-terminated and (b) Pd-terminated ribbons. Panels (c)-(d) are the same as panels (a)-(b), respectively, but with SOC. The Fermi level is set to zero. The edge states connect the bulk valence and conduction bands.}
	\label{fgr:fig-7}
\end{figure*}

The Fermi surfaces of monolayer $\beta$-Bi$_{2}$Pd
are shown in Fig. \ref{fgr:fig-5}.
Without SOC, two bands cross the Fermi level, forming three anisotropic pockets:
(i) one concave-square electron pocket at (0,0),
(ii) four water-drop shaped hole pockets along the $\Gamma$-M line,
and (iii) convex-square hole pockets around the corners.
With SOC, the four water-drop electron pockets become four round hole pockets,
while the convex-square hole pockets split into an inner and outer section
around the M point.

Interestingly, the Fermi surfaces of the hole pockets around neighboring M
points are almost parallel to each other.
This often gives rise to a Fermi surface nesting and phonon softening, which can contribute to the EPC.
The static electronic susceptibility $\chi'$ and the nesting factor $\chi''$ were calculated
by extracting the real part and imaginary part of the electronic susceptibility $\chi$
in the constant matrix element approximation\cite{Calandra2011,Zhuang2017}.
The $\chi'$ is given by
\begin{align}
\chi'(\mathbf{q})=\sum_{k}\frac{f(\varepsilon _{\mathbf{k}})-f(\varepsilon _{\mathbf{k+q}})}{\varepsilon _\mathbf{k}-\varepsilon _\mathbf{k+q}},
\end{align}
where $\varepsilon _{\mathbf{k}}$ and $\varepsilon _{\mathbf{k+q}}$
are band energies measured from the Fermi level at the wave vectors $\mathbf{k}$ and $\mathbf{k+q}$, respectively.
The $\chi''$ is defined by
\begin{align}
\lim_{\omega \rightarrow 0} \chi''(\mathbf{q},\omega )/\omega =\sum_{\mathbf{k}}\delta (\varepsilon _\mathbf{k}-\varepsilon _\mathbf{F})\delta (\varepsilon _\mathbf{k+q}-\varepsilon _\mathbf{F}),
\end{align}
where $\varepsilon _{F}$ is the Fermi energy.

Figure \ref{fgr:fig-6} shows the resulting $\chi'$ and $\chi''$,
calculated on a dense 200$\times$200$\times$1 $\textbf{k}$-mesh
using the Hamiltonian in the Wannier basis\cite{RevModPhys.84.1419}.
The inclusion of the SOC leads to a complete reconstruction of the susceptibility around the Fermi level,
showing the importance of SOC in this heavy element p-electron system.
In the case of $\chi'$ with SOC in Fig. \ref{fgr:fig-6}(c),
the strongest peaks occur along the $\Gamma$-M direction of the Brillouin zone.
The softening of the phonons in energy regions I and II mentioned above in the spectrum of
Fig. \ref{fgr:fig-4}(a) may thus stem from this structure in $\chi'$ structure.
In addition, $\chi''$ in Fig. \ref{fgr:fig-6}(d) shows strong peaks around $\Gamma$, which simply reflects
the density of states.
Other peak positions of $\chi''$ include
along the $\Gamma$-X direction.
This related to large contributions of $\lambda_{qn}$ in energy region II seen in Fig. \ref{fgr:fig-4}(a).

\subsection{Topological aspects}

We further explore how to achieve a topological superconducting phase in single-layer $\beta$-Bi$_{2}$Pd.
A full-gap feature in the bulk and topological surface states at its edge-terminated form
are the two key criteria to produce a topological superconductor,
which already appears in bulk $\beta$-Bi$_{2}$Pd\cite{Iwaya2017,LV2017852},
Au$_{2}$Pd\cite{PhysRevB.91.214517,Xing2016},
and PbTaSe$_{2}$\cite{Guan2016,Chang2016}, as shown by first-principles calculations and experimental measurements.
Monolayer $\beta$-Bi$_{2}$Pd having point group $\emph{D}$4$\emph{h}$,
is centrosymmetric and as mentioned
has a full energy gap throughout the whole BZ.
Using the tight-binding Wannier orbital based Hamiltonian\cite{RevModPhys.84.1419},
we performed edge-state calculations by the Green¡¯s-function approach\cite{Sancho1985}
for the corresponding semi-infinite systems
to explicitly verify edge states for monolayer $\beta$-Bi$_{2}$Pd.
The calculated results for monolayer $\beta$-Bi$_{2}$Pd without SOC,
presented in Figs. \ref{fgr:fig-7}(a) and \ref{fgr:fig-7}(b)
show that one gapless edge state connects the valence and conduction bands at $\rm \bar{X}$
point for Bi-terminated and Pd-terminated ribbons, respectively.
There are also additional edge states in Fig. \ref{fgr:fig-7},
owing to the existence of Dirac points in the band structure of Fig. \ref{fgr:fig-2}.
With SOC, the split bulk states harbor topological Dirac-like states
lying in the continuous energy gap around the $\rm \bar{X}$ point
[Figs. \ref{fgr:fig-7}(c) and \ref{fgr:fig-7}(d)].
This confirms the bulk-boundary correspondence between the bulk topology\cite{PhysRevB.76.045302,PhysRevLett.98.106803}
and the surface-state configurations, demonstrating that single-layer $\beta$-Bi$_{2}$Pd
indeed hosts edge states,
which is consistent with previous result\cite{Zhu2020PdBi}.

Although conventional 2D s-wave superconductors with time-reversal symmetry
theoretically are not topologically superconducting\cite{PhysRevB.81.134508},
the proximity effect in its topological surface states would induce the Majorana modes\cite{Jin2019MgB}.
Meanwhile, Masatoshi Sato proves that an s-wave superconductor can support Majorana fermion excitations if the pairing is realized in 2D Dirac fermions\cite{Sato2003}.
Fu et al. explicitly demonstrate that the Bogoliubov-de Gennes Hamiltonian of Dirac-type topological surface state,
couplings to a s-wave superconductor by proximity effect,
resembles a spinless $p$-wave superconductor which can realize Majorana zero modes\cite{PhysRevLett.100.096407}.
Usually, one could consider the Cooper pair tunneling into the surface states via the proximity effect
by introducing a superconducting gap to effective Hamiltonian
for single-layer $\beta$-Bi$_{2}$Pd. This is studied with a minimal three-band tight-binding model\cite{2008.06260}.
Then topological superconducting phase can be effectively induced via ideal substrates
depending on the parameters of the SOC and external field\cite{PhysRevLett.112.116404,Jin2019MgB}.

\section{Conclusions}

In summary, starting from the recently discovered topological superconductor $\beta$-Bi$_{2}$Pd,
we investigated the electronic structure and electron-phonon coupling of monolayer $\beta$-Bi$_{2}$Pd
by first-principles calculations.
We demonstrate that monolayer $\beta$-Bi$_{2}$Pd is an intrinsic
phonon-mediated superconductor.
Based on the Green¡¯s-function approach,
we have observed the Dirac-like edge states for its corresponding semi-infinite systems.
The superconductivity and topological
edge electrons could enable nontrivial superconducting pair
to possibly occur in $\beta$-Bi$_{2}$Pd.
Therefore, our studies provide a useful platform for realizing 1D topological superconductivity,
and the associated edge states and Majorana physics in single-layer $\beta$-Bi$_2$Pd.

\begin{acknowledgments}
We thank Hang Li from Henan University for useful discussions.
Peng-Fei Liu gratefully acknowledge financial support from the Guangdong Basic and Applied Basic Research Foundation
and the PhD Start-up Fund of the Natural Science Foundationof Guangdong Province of China (Grant No. 2018A0303100013).
Junrong Zhang acknowledges financial support from Natural Science Foundation of China (Grant No. 11675195).
The authors also thank the computational resources from the Supercomputer Centre of the China Spallation Neutron Source.
Work at the University of Missouri is supported by the Department of Energy, Basic Energy Sciences, Award DE-SC0019114.
\end{acknowledgments}


\begin{thebibliography}{76}%
\makeatletter
\providecommand \@ifxundefined [1]{%
 \@ifx{#1\undefined}
}%
\providecommand \@ifnum [1]{%
 \ifnum #1\expandafter \@firstoftwo
 \else \expandafter \@secondoftwo
 \fi
}%
\providecommand \@ifx [1]{%
 \ifx #1\expandafter \@firstoftwo
 \else \expandafter \@secondoftwo
 \fi
}%
\providecommand \natexlab [1]{#1}%
\providecommand \enquote  [1]{``#1''}%
\providecommand \bibnamefont  [1]{#1}%
\providecommand \bibfnamefont [1]{#1}%
\providecommand \citenamefont [1]{#1}%
\providecommand \href@noop [0]{\@secondoftwo}%
\providecommand \href [0]{\begingroup \@sanitize@url \@href}%
\providecommand \@href[1]{\@@startlink{#1}\@@href}%
\providecommand \@@href[1]{\endgroup#1\@@endlink}%
\providecommand \@sanitize@url [0]{\catcode `\\12\catcode `\$12\catcode
  `\&12\catcode `\#12\catcode `\^12\catcode `\_12\catcode `\%12\relax}%
\providecommand \@@startlink[1]{}%
\providecommand \@@endlink[0]{}%
\providecommand \url  [0]{\begingroup\@sanitize@url \@url }%
\providecommand \@url [1]{\endgroup\@href {#1}{\urlprefix }}%
\providecommand \urlprefix  [0]{URL }%
\providecommand \Eprint [0]{\href }%
\providecommand \doibase [0]{http://dx.doi.org/}%
\providecommand \selectlanguage [0]{\@gobble}%
\providecommand \bibinfo  [0]{\@secondoftwo}%
\providecommand \bibfield  [0]{\@secondoftwo}%
\providecommand \translation [1]{[#1]}%
\providecommand \BibitemOpen [0]{}%
\providecommand \bibitemStop [0]{}%
\providecommand \bibitemNoStop [0]{.\EOS\space}%
\providecommand \EOS [0]{\spacefactor3000\relax}%
\providecommand \BibitemShut  [1]{\csname bibitem#1\endcsname}%
\let\auto@bib@innerbib\@empty
\bibitem [{\citenamefont {Zhang}\ \emph {et~al.}(2010)\citenamefont {Zhang},
  \citenamefont {Cheng}, \citenamefont {Li}, \citenamefont {Sun}, \citenamefont
  {Wang}, \citenamefont {Zhu}, \citenamefont {He}, \citenamefont {Wang},
  \citenamefont {Ma}, \citenamefont {Chen}, \citenamefont {Wang}, \citenamefont
  {Liu}, \citenamefont {Lin}, \citenamefont {Jia},\ and\ \citenamefont
  {Xue}}]{Zhang2010}%
  \BibitemOpen
  \bibfield  {author} {\bibinfo {author} {\bibfnamefont {T.}~\bibnamefont
  {Zhang}}, \bibinfo {author} {\bibfnamefont {P.}~\bibnamefont {Cheng}},
  \bibinfo {author} {\bibfnamefont {W.-J.}\ \bibnamefont {Li}}, \bibinfo
  {author} {\bibfnamefont {Y.-J.}\ \bibnamefont {Sun}}, \bibinfo {author}
  {\bibfnamefont {G.}~\bibnamefont {Wang}}, \bibinfo {author} {\bibfnamefont
  {X.-G.}\ \bibnamefont {Zhu}}, \bibinfo {author} {\bibfnamefont
  {K.}~\bibnamefont {He}}, \bibinfo {author} {\bibfnamefont {L.}~\bibnamefont
  {Wang}}, \bibinfo {author} {\bibfnamefont {X.}~\bibnamefont {Ma}}, \bibinfo
  {author} {\bibfnamefont {X.}~\bibnamefont {Chen}}, \bibinfo {author}
  {\bibfnamefont {Y.}~\bibnamefont {Wang}}, \bibinfo {author} {\bibfnamefont
  {Y.}~\bibnamefont {Liu}}, \bibinfo {author} {\bibfnamefont {H.-Q.}\
  \bibnamefont {Lin}}, \bibinfo {author} {\bibfnamefont {J.-F.}\ \bibnamefont
  {Jia}}, \ and\ \bibinfo {author} {\bibfnamefont {Q.-K.}\ \bibnamefont
  {Xue}},\ }\href {\doibase 10.1038/nphys1499} {\bibfield  {journal} {\bibinfo
  {journal} {Nat. Phys.}\ }\textbf {\bibinfo {volume} {6}},\ \bibinfo {pages}
  {104} (\bibinfo {year} {2010})}\BibitemShut {NoStop}%
\bibitem [{\citenamefont {Eom}\ \emph {et~al.}(2006)\citenamefont {Eom},
  \citenamefont {Qin}, \citenamefont {Chou},\ and\ \citenamefont
  {Shih}}]{Eom2006}%
  \BibitemOpen
  \bibfield  {author} {\bibinfo {author} {\bibfnamefont {D.}~\bibnamefont
  {Eom}}, \bibinfo {author} {\bibfnamefont {S.}~\bibnamefont {Qin}}, \bibinfo
  {author} {\bibfnamefont {M.-Y.}\ \bibnamefont {Chou}}, \ and\ \bibinfo
  {author} {\bibfnamefont {C.~K.}\ \bibnamefont {Shih}},\ }\href {\doibase
  10.1103/PhysRevLett.96.027005} {\bibfield  {journal} {\bibinfo  {journal}
  {Phys. Rev. Lett.}\ }\textbf {\bibinfo {volume} {96}},\ \bibinfo {pages}
  {027005} (\bibinfo {year} {2006})}\BibitemShut {NoStop}%
\bibitem [{\citenamefont {He}\ \emph {et~al.}(2013)\citenamefont {He},
  \citenamefont {He}, \citenamefont {Zhang}, \citenamefont {Zhao},
  \citenamefont {Liu}, \citenamefont {Liu}, \citenamefont {Mou}, \citenamefont
  {Ou}, \citenamefont {Wang}, \citenamefont {Li}, \citenamefont {Wang},
  \citenamefont {Peng}, \citenamefont {Liu}, \citenamefont {Chen},
  \citenamefont {Yu}, \citenamefont {Liu}, \citenamefont {Dong}, \citenamefont
  {Zhang}, \citenamefont {Chen}, \citenamefont {Xu}, \citenamefont {Chen},
  \citenamefont {Ma}, \citenamefont {Xue},\ and\ \citenamefont
  {Zhou}}]{He2013}%
  \BibitemOpen
  \bibfield  {author} {\bibinfo {author} {\bibfnamefont {S.}~\bibnamefont
  {He}}, \bibinfo {author} {\bibfnamefont {J.}~\bibnamefont {He}}, \bibinfo
  {author} {\bibfnamefont {W.}~\bibnamefont {Zhang}}, \bibinfo {author}
  {\bibfnamefont {L.}~\bibnamefont {Zhao}}, \bibinfo {author} {\bibfnamefont
  {D.}~\bibnamefont {Liu}}, \bibinfo {author} {\bibfnamefont {X.}~\bibnamefont
  {Liu}}, \bibinfo {author} {\bibfnamefont {D.}~\bibnamefont {Mou}}, \bibinfo
  {author} {\bibfnamefont {Y.-B.}\ \bibnamefont {Ou}}, \bibinfo {author}
  {\bibfnamefont {Q.-Y.}\ \bibnamefont {Wang}}, \bibinfo {author}
  {\bibfnamefont {Z.}~\bibnamefont {Li}}, \bibinfo {author} {\bibfnamefont
  {L.}~\bibnamefont {Wang}}, \bibinfo {author} {\bibfnamefont {Y.}~\bibnamefont
  {Peng}}, \bibinfo {author} {\bibfnamefont {Y.}~\bibnamefont {Liu}}, \bibinfo
  {author} {\bibfnamefont {C.}~\bibnamefont {Chen}}, \bibinfo {author}
  {\bibfnamefont {L.}~\bibnamefont {Yu}}, \bibinfo {author} {\bibfnamefont
  {G.}~\bibnamefont {Liu}}, \bibinfo {author} {\bibfnamefont {X.}~\bibnamefont
  {Dong}}, \bibinfo {author} {\bibfnamefont {J.}~\bibnamefont {Zhang}},
  \bibinfo {author} {\bibfnamefont {C.}~\bibnamefont {Chen}}, \bibinfo {author}
  {\bibfnamefont {Z.}~\bibnamefont {Xu}}, \bibinfo {author} {\bibfnamefont
  {X.}~\bibnamefont {Chen}}, \bibinfo {author} {\bibfnamefont {X.}~\bibnamefont
  {Ma}}, \bibinfo {author} {\bibfnamefont {Q.}~\bibnamefont {Xue}}, \ and\
  \bibinfo {author} {\bibfnamefont {X.~J.}\ \bibnamefont {Zhou}},\ }\href
  {\doibase 10.1038/nmat3648} {\bibfield  {journal} {\bibinfo  {journal} {Nat.
  Mater.}\ }\textbf {\bibinfo {volume} {12}},\ \bibinfo {pages} {605} (\bibinfo
  {year} {2013})}\BibitemShut {NoStop}%
\bibitem [{\citenamefont {Ge}\ \emph {et~al.}(2014)\citenamefont {Ge},
  \citenamefont {Liu}, \citenamefont {Liu}, \citenamefont {Gao}, \citenamefont
  {Qian}, \citenamefont {Xue}, \citenamefont {Liu},\ and\ \citenamefont
  {Jia}}]{Ge2014}%
  \BibitemOpen
  \bibfield  {author} {\bibinfo {author} {\bibfnamefont {J.-F.}\ \bibnamefont
  {Ge}}, \bibinfo {author} {\bibfnamefont {Z.-L.}\ \bibnamefont {Liu}},
  \bibinfo {author} {\bibfnamefont {C.}~\bibnamefont {Liu}}, \bibinfo {author}
  {\bibfnamefont {C.-L.}\ \bibnamefont {Gao}}, \bibinfo {author} {\bibfnamefont
  {D.}~\bibnamefont {Qian}}, \bibinfo {author} {\bibfnamefont {Q.-K.}\
  \bibnamefont {Xue}}, \bibinfo {author} {\bibfnamefont {Y.}~\bibnamefont
  {Liu}}, \ and\ \bibinfo {author} {\bibfnamefont {J.-F.}\ \bibnamefont
  {Jia}},\ }\href {\doibase 10.1038/nmat4153} {\bibfield  {journal} {\bibinfo
  {journal} {Nat. Mater.}\ }\textbf {\bibinfo {volume} {14}},\ \bibinfo {pages}
  {285} (\bibinfo {year} {2014})}\BibitemShut {NoStop}%
\bibitem [{\citenamefont {Cao}\ \emph {et~al.}(2018)\citenamefont {Cao},
  \citenamefont {Fatemi}, \citenamefont {Fang}, \citenamefont {Watanabe},
  \citenamefont {Taniguchi}, \citenamefont {Kaxiras},\ and\ \citenamefont
  {Jarillo-Herrero}}]{cao2018}%
  \BibitemOpen
  \bibfield  {author} {\bibinfo {author} {\bibfnamefont {Y.}~\bibnamefont
  {Cao}}, \bibinfo {author} {\bibfnamefont {V.}~\bibnamefont {Fatemi}},
  \bibinfo {author} {\bibfnamefont {S.}~\bibnamefont {Fang}}, \bibinfo {author}
  {\bibfnamefont {K.}~\bibnamefont {Watanabe}}, \bibinfo {author}
  {\bibfnamefont {T.}~\bibnamefont {Taniguchi}}, \bibinfo {author}
  {\bibfnamefont {E.}~\bibnamefont {Kaxiras}}, \ and\ \bibinfo {author}
  {\bibfnamefont {P.}~\bibnamefont {Jarillo-Herrero}},\ }\href {\doibase
  10.1038/nature26160} {\bibfield  {journal} {\bibinfo  {journal} {Nature}\
  }\textbf {\bibinfo {volume} {556}},\ \bibinfo {pages} {43} (\bibinfo {year}
  {2018})}\BibitemShut {NoStop}%
\bibitem [{\citenamefont {Nam}\ \emph {et~al.}(2016)\citenamefont {Nam},
  \citenamefont {Chen}, \citenamefont {Liu}, \citenamefont {Kim}, \citenamefont
  {Zhang}, \citenamefont {Yong}, \citenamefont {Lemberger}, \citenamefont
  {Kratz}, \citenamefont {Kirtley}, \citenamefont {Moler}, \citenamefont
  {Adams}, \citenamefont {MacDonald},\ and\ \citenamefont {Shih}}]{nam}%
  \BibitemOpen
  \bibfield  {author} {\bibinfo {author} {\bibfnamefont {H.}~\bibnamefont
  {Nam}}, \bibinfo {author} {\bibfnamefont {H.}~\bibnamefont {Chen}}, \bibinfo
  {author} {\bibfnamefont {T.}~\bibnamefont {Liu}}, \bibinfo {author}
  {\bibfnamefont {J.}~\bibnamefont {Kim}}, \bibinfo {author} {\bibfnamefont
  {C.}~\bibnamefont {Zhang}}, \bibinfo {author} {\bibfnamefont
  {J.}~\bibnamefont {Yong}}, \bibinfo {author} {\bibfnamefont {T.~R.}\
  \bibnamefont {Lemberger}}, \bibinfo {author} {\bibfnamefont {P.~A.}\
  \bibnamefont {Kratz}}, \bibinfo {author} {\bibfnamefont {J.~R.}\ \bibnamefont
  {Kirtley}}, \bibinfo {author} {\bibfnamefont {K.}~\bibnamefont {Moler}},
  \bibinfo {author} {\bibfnamefont {P.~W.}\ \bibnamefont {Adams}}, \bibinfo
  {author} {\bibfnamefont {A.~H.}\ \bibnamefont {MacDonald}}, \ and\ \bibinfo
  {author} {\bibfnamefont {C.-K.}\ \bibnamefont {Shih}},\ }\href {\doibase
  10.1073/pnas.1611967113} {\bibfield  {journal} {\bibinfo  {journal} {Proc.
  Nat. Acad. Sci.}\ }\textbf {\bibinfo {volume} {113}},\ \bibinfo {pages}
  {10513} (\bibinfo {year} {2016})}\BibitemShut {NoStop}%
\bibitem [{\citenamefont {Yu}\ \emph {et~al.}(2019)\citenamefont {Yu},
  \citenamefont {Ma}, \citenamefont {Cai}, \citenamefont {Zhong}, \citenamefont
  {Ye}, \citenamefont {Shen}, \citenamefont {Gu}, \citenamefont {Chen},\ and\
  \citenamefont {Zhang}}]{yu2019}%
  \BibitemOpen
  \bibfield  {author} {\bibinfo {author} {\bibfnamefont {Y.}~\bibnamefont
  {Yu}}, \bibinfo {author} {\bibfnamefont {L.}~\bibnamefont {Ma}}, \bibinfo
  {author} {\bibfnamefont {P.}~\bibnamefont {Cai}}, \bibinfo {author}
  {\bibfnamefont {R.}~\bibnamefont {Zhong}}, \bibinfo {author} {\bibfnamefont
  {C.}~\bibnamefont {Ye}}, \bibinfo {author} {\bibfnamefont {J.}~\bibnamefont
  {Shen}}, \bibinfo {author} {\bibfnamefont {G.~D.}\ \bibnamefont {Gu}},
  \bibinfo {author} {\bibfnamefont {X.~H.}\ \bibnamefont {Chen}}, \ and\
  \bibinfo {author} {\bibfnamefont {Y.}~\bibnamefont {Zhang}},\ }\href
  {\doibase 10.1038/s41586-019-1718-x} {\bibfield  {journal} {\bibinfo
  {journal} {Nature}\ }\textbf {\bibinfo {volume} {575}},\ \bibinfo {pages}
  {156} (\bibinfo {year} {2019})}\BibitemShut {NoStop}%
\bibitem [{\citenamefont {Majorana}(1937)}]{Majorana1937}%
  \BibitemOpen
  \bibfield  {author} {\bibinfo {author} {\bibfnamefont {E.}~\bibnamefont
  {Majorana}},\ }\href {\doibase 10.1007/BF02961314} {\bibfield  {journal}
  {\bibinfo  {journal} {Il Nuovo Cimento}\ }\textbf {\bibinfo {volume} {14}},\
  \bibinfo {pages} {171} (\bibinfo {year} {1937})}\BibitemShut {NoStop}%
\bibitem [{\citenamefont {Nayak}\ \emph {et~al.}(2008)\citenamefont {Nayak},
  \citenamefont {Simon}, \citenamefont {Stern}, \citenamefont {Freedman},\ and\
  \citenamefont {Das~Sarma}}]{nayak}%
  \BibitemOpen
  \bibfield  {author} {\bibinfo {author} {\bibfnamefont {C.}~\bibnamefont
  {Nayak}}, \bibinfo {author} {\bibfnamefont {S.~H.}\ \bibnamefont {Simon}},
  \bibinfo {author} {\bibfnamefont {A.}~\bibnamefont {Stern}}, \bibinfo
  {author} {\bibfnamefont {M.}~\bibnamefont {Freedman}}, \ and\ \bibinfo
  {author} {\bibfnamefont {S.}~\bibnamefont {Das~Sarma}},\ }\href {\doibase
  10.1103/RevModPhys.80.1083} {\bibfield  {journal} {\bibinfo  {journal} {Rev.
  Mod. Phys.}\ }\textbf {\bibinfo {volume} {80}},\ \bibinfo {pages} {1083}
  (\bibinfo {year} {2008})}\BibitemShut {NoStop}%
\bibitem [{\citenamefont {Hor}\ \emph {et~al.}(2010)\citenamefont {Hor},
  \citenamefont {Williams}, \citenamefont {Checkelsky}, \citenamefont
  {Roushan}, \citenamefont {Seo}, \citenamefont {Xu}, \citenamefont
  {Zandbergen}, \citenamefont {Yazdani}, \citenamefont {Ong},\ and\
  \citenamefont {Cava}}]{Hor2010}%
  \BibitemOpen
  \bibfield  {author} {\bibinfo {author} {\bibfnamefont {Y.~S.}\ \bibnamefont
  {Hor}}, \bibinfo {author} {\bibfnamefont {A.~J.}\ \bibnamefont {Williams}},
  \bibinfo {author} {\bibfnamefont {J.~G.}\ \bibnamefont {Checkelsky}},
  \bibinfo {author} {\bibfnamefont {P.}~\bibnamefont {Roushan}}, \bibinfo
  {author} {\bibfnamefont {J.}~\bibnamefont {Seo}}, \bibinfo {author}
  {\bibfnamefont {Q.}~\bibnamefont {Xu}}, \bibinfo {author} {\bibfnamefont
  {H.~W.}\ \bibnamefont {Zandbergen}}, \bibinfo {author} {\bibfnamefont
  {A.}~\bibnamefont {Yazdani}}, \bibinfo {author} {\bibfnamefont {N.~P.}\
  \bibnamefont {Ong}}, \ and\ \bibinfo {author} {\bibfnamefont {R.~J.}\
  \bibnamefont {Cava}},\ }\href {\doibase 10.1103/PhysRevLett.104.057001}
  {\bibfield  {journal} {\bibinfo  {journal} {Phys. Rev. Lett.}\ }\textbf
  {\bibinfo {volume} {104}},\ \bibinfo {pages} {057001} (\bibinfo {year}
  {2010})}\BibitemShut {NoStop}%
\bibitem [{\citenamefont {Matano}\ \emph {et~al.}(2016)\citenamefont {Matano},
  \citenamefont {Kriener}, \citenamefont {Segawa}, \citenamefont {Ando},\ and\
  \citenamefont {Zheng}}]{Matano2016}%
  \BibitemOpen
  \bibfield  {author} {\bibinfo {author} {\bibfnamefont {K.}~\bibnamefont
  {Matano}}, \bibinfo {author} {\bibfnamefont {M.}~\bibnamefont {Kriener}},
  \bibinfo {author} {\bibfnamefont {K.}~\bibnamefont {Segawa}}, \bibinfo
  {author} {\bibfnamefont {Y.}~\bibnamefont {Ando}}, \ and\ \bibinfo {author}
  {\bibfnamefont {G.-Q.}\ \bibnamefont {Zheng}},\ }\href {\doibase
  10.1038/nphys3781} {\bibfield  {journal} {\bibinfo  {journal} {Nat. Phys.}\
  }\textbf {\bibinfo {volume} {12}},\ \bibinfo {pages} {852} (\bibinfo {year}
  {2016})}\BibitemShut {NoStop}%
\bibitem [{\citenamefont {Liu}\ \emph {et~al.}(2015)\citenamefont {Liu},
  \citenamefont {Yao}, \citenamefont {Shao}, \citenamefont {Zuo}, \citenamefont
  {Pi}, \citenamefont {Tan}, \citenamefont {Zhang},\ and\ \citenamefont
  {Zhang}}]{Liu2015}%
  \BibitemOpen
  \bibfield  {author} {\bibinfo {author} {\bibfnamefont {Z.}~\bibnamefont
  {Liu}}, \bibinfo {author} {\bibfnamefont {X.}~\bibnamefont {Yao}}, \bibinfo
  {author} {\bibfnamefont {J.}~\bibnamefont {Shao}}, \bibinfo {author}
  {\bibfnamefont {M.}~\bibnamefont {Zuo}}, \bibinfo {author} {\bibfnamefont
  {L.}~\bibnamefont {Pi}}, \bibinfo {author} {\bibfnamefont {S.}~\bibnamefont
  {Tan}}, \bibinfo {author} {\bibfnamefont {C.}~\bibnamefont {Zhang}}, \ and\
  \bibinfo {author} {\bibfnamefont {Y.}~\bibnamefont {Zhang}},\ }\href
  {\doibase 10.1021/jacs.5b06815} {\bibfield  {journal} {\bibinfo  {journal}
  {J. Am. Chem. Soc.}\ }\textbf {\bibinfo {volume} {137}},\ \bibinfo {pages}
  {10512} (\bibinfo {year} {2015})}\BibitemShut {NoStop}%
\bibitem [{\citenamefont {Xu}\ \emph {et~al.}(2015)\citenamefont {Xu},
  \citenamefont {Wang}, \citenamefont {Liu}, \citenamefont {Ge}, \citenamefont
  {Yang}, \citenamefont {Liu}, \citenamefont {Xu}, \citenamefont {Guan},
  \citenamefont {Gao}, \citenamefont {Qian}, \citenamefont {Liu}, \citenamefont
  {Wang}, \citenamefont {Zhang}, \citenamefont {Xue},\ and\ \citenamefont
  {Jia}}]{Xu2015}%
  \BibitemOpen
  \bibfield  {author} {\bibinfo {author} {\bibfnamefont {J.-P.}\ \bibnamefont
  {Xu}}, \bibinfo {author} {\bibfnamefont {M.-X.}\ \bibnamefont {Wang}},
  \bibinfo {author} {\bibfnamefont {Z.~L.}\ \bibnamefont {Liu}}, \bibinfo
  {author} {\bibfnamefont {J.-F.}\ \bibnamefont {Ge}}, \bibinfo {author}
  {\bibfnamefont {X.}~\bibnamefont {Yang}}, \bibinfo {author} {\bibfnamefont
  {C.}~\bibnamefont {Liu}}, \bibinfo {author} {\bibfnamefont {Z.~A.}\
  \bibnamefont {Xu}}, \bibinfo {author} {\bibfnamefont {D.}~\bibnamefont
  {Guan}}, \bibinfo {author} {\bibfnamefont {C.~L.}\ \bibnamefont {Gao}},
  \bibinfo {author} {\bibfnamefont {D.}~\bibnamefont {Qian}}, \bibinfo {author}
  {\bibfnamefont {Y.}~\bibnamefont {Liu}}, \bibinfo {author} {\bibfnamefont
  {Q.-H.}\ \bibnamefont {Wang}}, \bibinfo {author} {\bibfnamefont {F.-C.}\
  \bibnamefont {Zhang}}, \bibinfo {author} {\bibfnamefont {Q.-K.}\ \bibnamefont
  {Xue}}, \ and\ \bibinfo {author} {\bibfnamefont {J.-F.}\ \bibnamefont
  {Jia}},\ }\href {\doibase 10.1103/PhysRevLett.114.017001} {\bibfield
  {journal} {\bibinfo  {journal} {Phys. Rev. Lett.}\ }\textbf {\bibinfo
  {volume} {114}},\ \bibinfo {pages} {017001} (\bibinfo {year}
  {2015})}\BibitemShut {NoStop}%
\bibitem [{\citenamefont {Sun}\ \emph {et~al.}(2016)\citenamefont {Sun},
  \citenamefont {Zhang}, \citenamefont {Hu}, \citenamefont {Li}, \citenamefont
  {Wang}, \citenamefont {Ma}, \citenamefont {Xu}, \citenamefont {Gao},
  \citenamefont {Guan}, \citenamefont {Li}, \citenamefont {Liu}, \citenamefont
  {Qian}, \citenamefont {Zhou}, \citenamefont {Fu}, \citenamefont {Li},
  \citenamefont {Zhang},\ and\ \citenamefont {Jia}}]{Sun2016}%
  \BibitemOpen
  \bibfield  {author} {\bibinfo {author} {\bibfnamefont {H.-H.}\ \bibnamefont
  {Sun}}, \bibinfo {author} {\bibfnamefont {K.-W.}\ \bibnamefont {Zhang}},
  \bibinfo {author} {\bibfnamefont {L.-H.}\ \bibnamefont {Hu}}, \bibinfo
  {author} {\bibfnamefont {C.}~\bibnamefont {Li}}, \bibinfo {author}
  {\bibfnamefont {G.-Y.}\ \bibnamefont {Wang}}, \bibinfo {author}
  {\bibfnamefont {H.-Y.}\ \bibnamefont {Ma}}, \bibinfo {author} {\bibfnamefont
  {Z.-A.}\ \bibnamefont {Xu}}, \bibinfo {author} {\bibfnamefont {C.-L.}\
  \bibnamefont {Gao}}, \bibinfo {author} {\bibfnamefont {D.-D.}\ \bibnamefont
  {Guan}}, \bibinfo {author} {\bibfnamefont {Y.-Y.}\ \bibnamefont {Li}},
  \bibinfo {author} {\bibfnamefont {C.}~\bibnamefont {Liu}}, \bibinfo {author}
  {\bibfnamefont {D.}~\bibnamefont {Qian}}, \bibinfo {author} {\bibfnamefont
  {Y.}~\bibnamefont {Zhou}}, \bibinfo {author} {\bibfnamefont {L.}~\bibnamefont
  {Fu}}, \bibinfo {author} {\bibfnamefont {S.-C.}\ \bibnamefont {Li}}, \bibinfo
  {author} {\bibfnamefont {F.-C.}\ \bibnamefont {Zhang}}, \ and\ \bibinfo
  {author} {\bibfnamefont {J.-F.}\ \bibnamefont {Jia}},\ }\href {\doibase
  10.1103/PhysRevLett.116.257003} {\bibfield  {journal} {\bibinfo  {journal}
  {Phys. Rev. Lett.}\ }\textbf {\bibinfo {volume} {116}},\ \bibinfo {pages}
  {257003} (\bibinfo {year} {2016})}\BibitemShut {NoStop}%
\bibitem [{\citenamefont {Zareapour}\ \emph {et~al.}(2012)\citenamefont
  {Zareapour}, \citenamefont {Hayat}, \citenamefont {Zhao}, \citenamefont
  {Kreshchuk}, \citenamefont {Jain}, \citenamefont {Kwok}, \citenamefont {Lee},
  \citenamefont {Cheong}, \citenamefont {Xu}, \citenamefont {Yang},
  \citenamefont {Gu}, \citenamefont {Jia}, \citenamefont {Cava},\ and\
  \citenamefont {Burch}}]{Zareapour2012}%
  \BibitemOpen
  \bibfield  {author} {\bibinfo {author} {\bibfnamefont {P.}~\bibnamefont
  {Zareapour}}, \bibinfo {author} {\bibfnamefont {A.}~\bibnamefont {Hayat}},
  \bibinfo {author} {\bibfnamefont {S.~Y.~F.}\ \bibnamefont {Zhao}}, \bibinfo
  {author} {\bibfnamefont {M.}~\bibnamefont {Kreshchuk}}, \bibinfo {author}
  {\bibfnamefont {A.}~\bibnamefont {Jain}}, \bibinfo {author} {\bibfnamefont
  {D.~C.}\ \bibnamefont {Kwok}}, \bibinfo {author} {\bibfnamefont
  {N.}~\bibnamefont {Lee}}, \bibinfo {author} {\bibfnamefont {S.-W.}\
  \bibnamefont {Cheong}}, \bibinfo {author} {\bibfnamefont {Z.}~\bibnamefont
  {Xu}}, \bibinfo {author} {\bibfnamefont {A.}~\bibnamefont {Yang}}, \bibinfo
  {author} {\bibfnamefont {G.}~\bibnamefont {Gu}}, \bibinfo {author}
  {\bibfnamefont {S.}~\bibnamefont {Jia}}, \bibinfo {author} {\bibfnamefont
  {R.~J.}\ \bibnamefont {Cava}}, \ and\ \bibinfo {author} {\bibfnamefont
  {K.~S.}\ \bibnamefont {Burch}},\ }\href {\doibase 10.1038/ncomms2042}
  {\bibfield  {journal} {\bibinfo  {journal} {Nat. Commun.}\ }\textbf {\bibinfo
  {volume} {3}},\ \bibinfo {pages} {1} (\bibinfo {year} {2012})}\BibitemShut
  {NoStop}%
\bibitem [{\citenamefont {Nadj-Perge}\ \emph {et~al.}(2014)\citenamefont
  {Nadj-Perge}, \citenamefont {Drozdov}, \citenamefont {Li}, \citenamefont
  {Chen}, \citenamefont {Jeon}, \citenamefont {Seo}, \citenamefont {MacDonald},
  \citenamefont {Bernevig},\ and\ \citenamefont {Yazdani}}]{NadjPerge2014}%
  \BibitemOpen
  \bibfield  {author} {\bibinfo {author} {\bibfnamefont {S.}~\bibnamefont
  {Nadj-Perge}}, \bibinfo {author} {\bibfnamefont {I.~K.}\ \bibnamefont
  {Drozdov}}, \bibinfo {author} {\bibfnamefont {J.}~\bibnamefont {Li}},
  \bibinfo {author} {\bibfnamefont {H.}~\bibnamefont {Chen}}, \bibinfo {author}
  {\bibfnamefont {S.}~\bibnamefont {Jeon}}, \bibinfo {author} {\bibfnamefont
  {J.}~\bibnamefont {Seo}}, \bibinfo {author} {\bibfnamefont {A.~H.}\
  \bibnamefont {MacDonald}}, \bibinfo {author} {\bibfnamefont {B.~A.}\
  \bibnamefont {Bernevig}}, \ and\ \bibinfo {author} {\bibfnamefont
  {A.}~\bibnamefont {Yazdani}},\ }\href {\doibase 10.1126/science.1259327}
  {\bibfield  {journal} {\bibinfo  {journal} {Science}\ }\textbf {\bibinfo
  {volume} {346}},\ \bibinfo {pages} {602} (\bibinfo {year}
  {2014})}\BibitemShut {NoStop}%
\bibitem [{\citenamefont {He}\ \emph {et~al.}(2017)\citenamefont {He},
  \citenamefont {Pan}, \citenamefont {Stern}, \citenamefont {Burks},
  \citenamefont {Che}, \citenamefont {Yin}, \citenamefont {Wang}, \citenamefont
  {Lian}, \citenamefont {Zhou}, \citenamefont {Choi}, \citenamefont {Murata},
  \citenamefont {Kou}, \citenamefont {Chen}, \citenamefont {Nie}, \citenamefont
  {Shao}, \citenamefont {Fan}, \citenamefont {Zhang}, \citenamefont {Liu},
  \citenamefont {Xia},\ and\ \citenamefont {Wang}}]{He2017}%
  \BibitemOpen
  \bibfield  {author} {\bibinfo {author} {\bibfnamefont {Q.~L.}\ \bibnamefont
  {He}}, \bibinfo {author} {\bibfnamefont {L.}~\bibnamefont {Pan}}, \bibinfo
  {author} {\bibfnamefont {A.~L.}\ \bibnamefont {Stern}}, \bibinfo {author}
  {\bibfnamefont {E.~C.}\ \bibnamefont {Burks}}, \bibinfo {author}
  {\bibfnamefont {X.}~\bibnamefont {Che}}, \bibinfo {author} {\bibfnamefont
  {G.}~\bibnamefont {Yin}}, \bibinfo {author} {\bibfnamefont {J.}~\bibnamefont
  {Wang}}, \bibinfo {author} {\bibfnamefont {B.}~\bibnamefont {Lian}}, \bibinfo
  {author} {\bibfnamefont {Q.}~\bibnamefont {Zhou}}, \bibinfo {author}
  {\bibfnamefont {E.~S.}\ \bibnamefont {Choi}}, \bibinfo {author}
  {\bibfnamefont {K.}~\bibnamefont {Murata}}, \bibinfo {author} {\bibfnamefont
  {X.}~\bibnamefont {Kou}}, \bibinfo {author} {\bibfnamefont {Z.}~\bibnamefont
  {Chen}}, \bibinfo {author} {\bibfnamefont {T.}~\bibnamefont {Nie}}, \bibinfo
  {author} {\bibfnamefont {Q.}~\bibnamefont {Shao}}, \bibinfo {author}
  {\bibfnamefont {Y.}~\bibnamefont {Fan}}, \bibinfo {author} {\bibfnamefont
  {S.-C.}\ \bibnamefont {Zhang}}, \bibinfo {author} {\bibfnamefont
  {K.}~\bibnamefont {Liu}}, \bibinfo {author} {\bibfnamefont {J.}~\bibnamefont
  {Xia}}, \ and\ \bibinfo {author} {\bibfnamefont {K.~L.}\ \bibnamefont
  {Wang}},\ }\href {\doibase 10.1126/science.aag2792} {\bibfield  {journal}
  {\bibinfo  {journal} {Science}\ }\textbf {\bibinfo {volume} {357}},\ \bibinfo
  {pages} {294} (\bibinfo {year} {2017})}\BibitemShut {NoStop}%
\bibitem [{\citenamefont {Kirshenbaum}\ \emph {et~al.}(2013)\citenamefont
  {Kirshenbaum}, \citenamefont {Syers}, \citenamefont {Hope}, \citenamefont
  {Butch}, \citenamefont {Jeffries}, \citenamefont {Weir}, \citenamefont
  {Hamlin}, \citenamefont {Maple}, \citenamefont {Vohra},\ and\ \citenamefont
  {Paglione}}]{PhysRevLett.111.087001}%
  \BibitemOpen
  \bibfield  {author} {\bibinfo {author} {\bibfnamefont {K.}~\bibnamefont
  {Kirshenbaum}}, \bibinfo {author} {\bibfnamefont {P.~S.}\ \bibnamefont
  {Syers}}, \bibinfo {author} {\bibfnamefont {A.~P.}\ \bibnamefont {Hope}},
  \bibinfo {author} {\bibfnamefont {N.~P.}\ \bibnamefont {Butch}}, \bibinfo
  {author} {\bibfnamefont {J.~R.}\ \bibnamefont {Jeffries}}, \bibinfo {author}
  {\bibfnamefont {S.~T.}\ \bibnamefont {Weir}}, \bibinfo {author}
  {\bibfnamefont {J.~J.}\ \bibnamefont {Hamlin}}, \bibinfo {author}
  {\bibfnamefont {M.~B.}\ \bibnamefont {Maple}}, \bibinfo {author}
  {\bibfnamefont {Y.~K.}\ \bibnamefont {Vohra}}, \ and\ \bibinfo {author}
  {\bibfnamefont {J.}~\bibnamefont {Paglione}},\ }\href {\doibase
  10.1103/PhysRevLett.111.087001} {\bibfield  {journal} {\bibinfo  {journal}
  {Phys. Rev. Lett.}\ }\textbf {\bibinfo {volume} {111}},\ \bibinfo {pages}
  {087001} (\bibinfo {year} {2013})}\BibitemShut {NoStop}%
\bibitem [{\citenamefont {Zhao}\ \emph {et~al.}(2015)\citenamefont {Zhao},
  \citenamefont {Deng}, \citenamefont {Korzhovska}, \citenamefont
  {Begliarbekov}, \citenamefont {Chen}, \citenamefont {Andrade}, \citenamefont
  {Rosenthal}, \citenamefont {Pasupathy}, \citenamefont {Oganesyan},\ and\
  \citenamefont {Krusin-Elbaum}}]{Zhao2015}%
  \BibitemOpen
  \bibfield  {author} {\bibinfo {author} {\bibfnamefont {L.}~\bibnamefont
  {Zhao}}, \bibinfo {author} {\bibfnamefont {H.}~\bibnamefont {Deng}}, \bibinfo
  {author} {\bibfnamefont {I.}~\bibnamefont {Korzhovska}}, \bibinfo {author}
  {\bibfnamefont {M.}~\bibnamefont {Begliarbekov}}, \bibinfo {author}
  {\bibfnamefont {Z.}~\bibnamefont {Chen}}, \bibinfo {author} {\bibfnamefont
  {E.}~\bibnamefont {Andrade}}, \bibinfo {author} {\bibfnamefont
  {E.}~\bibnamefont {Rosenthal}}, \bibinfo {author} {\bibfnamefont
  {A.}~\bibnamefont {Pasupathy}}, \bibinfo {author} {\bibfnamefont
  {V.}~\bibnamefont {Oganesyan}}, \ and\ \bibinfo {author} {\bibfnamefont
  {L.}~\bibnamefont {Krusin-Elbaum}},\ }\href {\doibase 10.1038/ncomms9279}
  {\bibfield  {journal} {\bibinfo  {journal} {Nat. Commun.}\ }\textbf {\bibinfo
  {volume} {6}},\ \bibinfo {pages} {1} (\bibinfo {year} {2015})}\BibitemShut
  {NoStop}%
\bibitem [{\citenamefont {Olego}(1989)}]{PhysRevB.39.12743}%
  \BibitemOpen
  \bibfield  {author} {\bibinfo {author} {\bibfnamefont {D.~J.}\ \bibnamefont
  {Olego}},\ }\href {\doibase 10.1103/PhysRevB.39.12743} {\bibfield  {journal}
  {\bibinfo  {journal} {Phys. Rev. B}\ }\textbf {\bibinfo {volume} {39}},\
  \bibinfo {pages} {12743} (\bibinfo {year} {1989})}\BibitemShut {NoStop}%
\bibitem [{\citenamefont {Mahendiran}\ \emph {et~al.}(2001)\citenamefont
  {Mahendiran}, \citenamefont {Raveau}, \citenamefont {Hervieu}, \citenamefont
  {Michel},\ and\ \citenamefont {Maignan}}]{PhysRevB.64.064424}%
  \BibitemOpen
  \bibfield  {author} {\bibinfo {author} {\bibfnamefont {R.}~\bibnamefont
  {Mahendiran}}, \bibinfo {author} {\bibfnamefont {B.}~\bibnamefont {Raveau}},
  \bibinfo {author} {\bibfnamefont {M.}~\bibnamefont {Hervieu}}, \bibinfo
  {author} {\bibfnamefont {C.}~\bibnamefont {Michel}}, \ and\ \bibinfo {author}
  {\bibfnamefont {A.}~\bibnamefont {Maignan}},\ }\href {\doibase
  10.1103/PhysRevB.64.064424} {\bibfield  {journal} {\bibinfo  {journal} {Phys.
  Rev. B}\ }\textbf {\bibinfo {volume} {64}},\ \bibinfo {pages} {064424}
  (\bibinfo {year} {2001})}\BibitemShut {NoStop}%
\bibitem [{\citenamefont {Sakano}\ \emph {et~al.}(2015)\citenamefont {Sakano},
  \citenamefont {Okawa}, \citenamefont {Kanou}, \citenamefont {Sanjo},
  \citenamefont {Okuda}, \citenamefont {Sasagawa},\ and\ \citenamefont
  {Ishizaka}}]{Sakano2015}%
  \BibitemOpen
  \bibfield  {author} {\bibinfo {author} {\bibfnamefont {M.}~\bibnamefont
  {Sakano}}, \bibinfo {author} {\bibfnamefont {K.}~\bibnamefont {Okawa}},
  \bibinfo {author} {\bibfnamefont {M.}~\bibnamefont {Kanou}}, \bibinfo
  {author} {\bibfnamefont {H.}~\bibnamefont {Sanjo}}, \bibinfo {author}
  {\bibfnamefont {T.}~\bibnamefont {Okuda}}, \bibinfo {author} {\bibfnamefont
  {T.}~\bibnamefont {Sasagawa}}, \ and\ \bibinfo {author} {\bibfnamefont
  {K.}~\bibnamefont {Ishizaka}},\ }\href {\doibase 10.1038/ncomms9595}
  {\bibfield  {journal} {\bibinfo  {journal} {Nat. Commun.}\ }\textbf {\bibinfo
  {volume} {6}},\ \bibinfo {pages} {1} (\bibinfo {year} {2015})}\BibitemShut
  {NoStop}%
\bibitem [{\citenamefont {Iwaya}\ \emph {et~al.}(2017)\citenamefont {Iwaya},
  \citenamefont {Kohsaka}, \citenamefont {Okawa}, \citenamefont {Machida},
  \citenamefont {Bahramy}, \citenamefont {Hanaguri},\ and\ \citenamefont
  {Sasagawa}}]{Iwaya2017}%
  \BibitemOpen
  \bibfield  {author} {\bibinfo {author} {\bibfnamefont {K.}~\bibnamefont
  {Iwaya}}, \bibinfo {author} {\bibfnamefont {Y.}~\bibnamefont {Kohsaka}},
  \bibinfo {author} {\bibfnamefont {K.}~\bibnamefont {Okawa}}, \bibinfo
  {author} {\bibfnamefont {T.}~\bibnamefont {Machida}}, \bibinfo {author}
  {\bibfnamefont {M.}~\bibnamefont {Bahramy}}, \bibinfo {author} {\bibfnamefont
  {T.}~\bibnamefont {Hanaguri}}, \ and\ \bibinfo {author} {\bibfnamefont
  {T.}~\bibnamefont {Sasagawa}},\ }\href {\doibase 10.1038/s41467-017-01209-9}
  {\bibfield  {journal} {\bibinfo  {journal} {Nat. Commun.}\ }\textbf {\bibinfo
  {volume} {8}},\ \bibinfo {pages} {1} (\bibinfo {year} {2017})}\BibitemShut
  {NoStop}%
\bibitem [{\citenamefont {Guan}\ \emph {et~al.}(2016)\citenamefont {Guan},
  \citenamefont {Chen}, \citenamefont {Chu}, \citenamefont {Sankar},
  \citenamefont {Chou}, \citenamefont {Jeng}, \citenamefont {Chang},\ and\
  \citenamefont {Chuang}}]{Guan2016}%
  \BibitemOpen
  \bibfield  {author} {\bibinfo {author} {\bibfnamefont {S.-Y.}\ \bibnamefont
  {Guan}}, \bibinfo {author} {\bibfnamefont {P.-J.}\ \bibnamefont {Chen}},
  \bibinfo {author} {\bibfnamefont {M.-W.}\ \bibnamefont {Chu}}, \bibinfo
  {author} {\bibfnamefont {R.}~\bibnamefont {Sankar}}, \bibinfo {author}
  {\bibfnamefont {F.}~\bibnamefont {Chou}}, \bibinfo {author} {\bibfnamefont
  {H.-T.}\ \bibnamefont {Jeng}}, \bibinfo {author} {\bibfnamefont {C.-S.}\
  \bibnamefont {Chang}}, \ and\ \bibinfo {author} {\bibfnamefont {T.-M.}\
  \bibnamefont {Chuang}},\ }\href {\doibase 10.1126/sciadv.1600894} {\bibfield
  {journal} {\bibinfo  {journal} {Sci. Adv.}\ }\textbf {\bibinfo {volume}
  {2}},\ \bibinfo {pages} {e1600894} (\bibinfo {year} {2016})}\BibitemShut
  {NoStop}%
\bibitem [{\citenamefont {Wang}\ \emph {et~al.}(2020)\citenamefont {Wang},
  \citenamefont {Kim}, \citenamefont {Liu}, \citenamefont {Cevallos},
  \citenamefont {Cava},\ and\ \citenamefont {Ong}}]{Wang2020}%
  \BibitemOpen
  \bibfield  {author} {\bibinfo {author} {\bibfnamefont {W.}~\bibnamefont
  {Wang}}, \bibinfo {author} {\bibfnamefont {S.}~\bibnamefont {Kim}}, \bibinfo
  {author} {\bibfnamefont {M.}~\bibnamefont {Liu}}, \bibinfo {author}
  {\bibfnamefont {F.~A.}\ \bibnamefont {Cevallos}}, \bibinfo {author}
  {\bibfnamefont {R.~J.}\ \bibnamefont {Cava}}, \ and\ \bibinfo {author}
  {\bibfnamefont {N.~P.}\ \bibnamefont {Ong}},\ }\href {\doibase
  10.1126/science.aaw9270} {\bibfield  {journal} {\bibinfo  {journal}
  {Science}\ }\textbf {\bibinfo {volume} {368}},\ \bibinfo {pages} {534}
  (\bibinfo {year} {2020})}\BibitemShut {NoStop}%
\bibitem [{\citenamefont {Yuan}\ \emph {et~al.}(2019)\citenamefont {Yuan},
  \citenamefont {Pan}, \citenamefont {Wang}, \citenamefont {Fang},
  \citenamefont {Song}, \citenamefont {Wang}, \citenamefont {He}, \citenamefont
  {Ma}, \citenamefont {Zhang}, \citenamefont {Huang}, \citenamefont {Li},\ and\
  \citenamefont {Xue}}]{Yuan2019}%
  \BibitemOpen
  \bibfield  {author} {\bibinfo {author} {\bibfnamefont {Y.}~\bibnamefont
  {Yuan}}, \bibinfo {author} {\bibfnamefont {J.}~\bibnamefont {Pan}}, \bibinfo
  {author} {\bibfnamefont {X.}~\bibnamefont {Wang}}, \bibinfo {author}
  {\bibfnamefont {Y.}~\bibnamefont {Fang}}, \bibinfo {author} {\bibfnamefont
  {C.}~\bibnamefont {Song}}, \bibinfo {author} {\bibfnamefont {L.}~\bibnamefont
  {Wang}}, \bibinfo {author} {\bibfnamefont {K.}~\bibnamefont {He}}, \bibinfo
  {author} {\bibfnamefont {X.}~\bibnamefont {Ma}}, \bibinfo {author}
  {\bibfnamefont {H.}~\bibnamefont {Zhang}}, \bibinfo {author} {\bibfnamefont
  {F.}~\bibnamefont {Huang}}, \bibinfo {author} {\bibfnamefont
  {W.}~\bibnamefont {Li}}, \ and\ \bibinfo {author} {\bibfnamefont {Q.-K.}\
  \bibnamefont {Xue}},\ }\href {\doibase 10.1038/s41567-019-0576-7} {\bibfield
  {journal} {\bibinfo  {journal} {Nat. Phys.}\ }\textbf {\bibinfo {volume}
  {15}},\ \bibinfo {pages} {1046} (\bibinfo {year} {2019})}\BibitemShut
  {NoStop}%
\bibitem [{\citenamefont {Fang}\ \emph {et~al.}(2019)\citenamefont {Fang},
  \citenamefont {Pan}, \citenamefont {Zhang}, \citenamefont {Wang},
  \citenamefont {Hirose}, \citenamefont {Terashima}, \citenamefont {Uji},
  \citenamefont {Yuan}, \citenamefont {Li}, \citenamefont {Tian}, \citenamefont
  {Xue}, \citenamefont {Ma}, \citenamefont {Zhao}, \citenamefont {Xue},
  \citenamefont {Mu}, \citenamefont {Zhang},\ and\ \citenamefont
  {Huang}}]{fang2019discovery}%
  \BibitemOpen
  \bibfield  {author} {\bibinfo {author} {\bibfnamefont {Y.}~\bibnamefont
  {Fang}}, \bibinfo {author} {\bibfnamefont {J.}~\bibnamefont {Pan}}, \bibinfo
  {author} {\bibfnamefont {D.}~\bibnamefont {Zhang}}, \bibinfo {author}
  {\bibfnamefont {D.}~\bibnamefont {Wang}}, \bibinfo {author} {\bibfnamefont
  {H.~T.}\ \bibnamefont {Hirose}}, \bibinfo {author} {\bibfnamefont
  {T.}~\bibnamefont {Terashima}}, \bibinfo {author} {\bibfnamefont
  {S.}~\bibnamefont {Uji}}, \bibinfo {author} {\bibfnamefont {Y.}~\bibnamefont
  {Yuan}}, \bibinfo {author} {\bibfnamefont {W.}~\bibnamefont {Li}}, \bibinfo
  {author} {\bibfnamefont {Z.}~\bibnamefont {Tian}}, \bibinfo {author}
  {\bibfnamefont {J.}~\bibnamefont {Xue}}, \bibinfo {author} {\bibfnamefont
  {Y.}~\bibnamefont {Ma}}, \bibinfo {author} {\bibfnamefont {W.}~\bibnamefont
  {Zhao}}, \bibinfo {author} {\bibfnamefont {Q.}~\bibnamefont {Xue}}, \bibinfo
  {author} {\bibfnamefont {G.}~\bibnamefont {Mu}}, \bibinfo {author}
  {\bibfnamefont {H.}~\bibnamefont {Zhang}}, \ and\ \bibinfo {author}
  {\bibfnamefont {F.}~\bibnamefont {Huang}},\ }\href {\doibase
  10.1002/adma.201901942} {\bibfield  {journal} {\bibinfo  {journal} {Adv.
  Mater.}\ }\textbf {\bibinfo {volume} {31}},\ \bibinfo {pages} {1901942}
  (\bibinfo {year} {2019})}\BibitemShut {NoStop}%
\bibitem [{\citenamefont {Guguchia}\ \emph {et~al.}(2017)\citenamefont
  {Guguchia}, \citenamefont {von Rohr}, \citenamefont {Shermadini},
  \citenamefont {Lee}, \citenamefont {Banerjee}, \citenamefont {Wieteska},
  \citenamefont {Marianetti}, \citenamefont {Frandsen}, \citenamefont
  {Luetkens}, \citenamefont {Gong}, \citenamefont {Cheung}, \citenamefont
  {Baines}, \citenamefont {Shengelaya}, \citenamefont {Taniashvili},
  \citenamefont {Pasupathy}, \citenamefont {Morenzoni}, \citenamefont
  {Billinge}, \citenamefont {Amato}, \citenamefont {Cava}, \citenamefont
  {Khasanov},\ and\ \citenamefont {Uemura}}]{guguchia2017signatures}%
  \BibitemOpen
  \bibfield  {author} {\bibinfo {author} {\bibfnamefont {Z.}~\bibnamefont
  {Guguchia}}, \bibinfo {author} {\bibfnamefont {F.}~\bibnamefont {von Rohr}},
  \bibinfo {author} {\bibfnamefont {Z.}~\bibnamefont {Shermadini}}, \bibinfo
  {author} {\bibfnamefont {A.~T.}\ \bibnamefont {Lee}}, \bibinfo {author}
  {\bibfnamefont {S.}~\bibnamefont {Banerjee}}, \bibinfo {author}
  {\bibfnamefont {A.~R.}\ \bibnamefont {Wieteska}}, \bibinfo {author}
  {\bibfnamefont {C.~A.}\ \bibnamefont {Marianetti}}, \bibinfo {author}
  {\bibfnamefont {B.~A.}\ \bibnamefont {Frandsen}}, \bibinfo {author}
  {\bibfnamefont {H.}~\bibnamefont {Luetkens}}, \bibinfo {author}
  {\bibfnamefont {Z.}~\bibnamefont {Gong}}, \bibinfo {author} {\bibfnamefont
  {S.~C.}\ \bibnamefont {Cheung}}, \bibinfo {author} {\bibfnamefont
  {C.}~\bibnamefont {Baines}}, \bibinfo {author} {\bibfnamefont
  {A.}~\bibnamefont {Shengelaya}}, \bibinfo {author} {\bibfnamefont
  {G.}~\bibnamefont {Taniashvili}}, \bibinfo {author} {\bibfnamefont {A.~N.}\
  \bibnamefont {Pasupathy}}, \bibinfo {author} {\bibfnamefont {E.}~\bibnamefont
  {Morenzoni}}, \bibinfo {author} {\bibfnamefont {S.~J.~L.}\ \bibnamefont
  {Billinge}}, \bibinfo {author} {\bibfnamefont {A.}~\bibnamefont {Amato}},
  \bibinfo {author} {\bibfnamefont {R.~J.}\ \bibnamefont {Cava}}, \bibinfo
  {author} {\bibfnamefont {R.}~\bibnamefont {Khasanov}}, \ and\ \bibinfo
  {author} {\bibfnamefont {Y.~J.}\ \bibnamefont {Uemura}},\ }\href {\doibase
  10.1038/s41467-017-01066-6} {\bibfield  {journal} {\bibinfo  {journal} {Nat.
  Commun.}\ }\textbf {\bibinfo {volume} {8}},\ \bibinfo {pages} {1} (\bibinfo
  {year} {2017})}\BibitemShut {NoStop}%
\bibitem [{\citenamefont {Sun}\ \emph {et~al.}(2015)\citenamefont {Sun},
  \citenamefont {Enayat}, \citenamefont {Maldonado}, \citenamefont {Lithgow},
  \citenamefont {Yelland}, \citenamefont {Peets}, \citenamefont {Yaresko},
  \citenamefont {Schnyder},\ and\ \citenamefont {Wahl}}]{Sun2015}%
  \BibitemOpen
  \bibfield  {author} {\bibinfo {author} {\bibfnamefont {Z.}~\bibnamefont
  {Sun}}, \bibinfo {author} {\bibfnamefont {M.}~\bibnamefont {Enayat}},
  \bibinfo {author} {\bibfnamefont {A.}~\bibnamefont {Maldonado}}, \bibinfo
  {author} {\bibfnamefont {C.}~\bibnamefont {Lithgow}}, \bibinfo {author}
  {\bibfnamefont {E.}~\bibnamefont {Yelland}}, \bibinfo {author} {\bibfnamefont
  {D.~C.}\ \bibnamefont {Peets}}, \bibinfo {author} {\bibfnamefont
  {A.}~\bibnamefont {Yaresko}}, \bibinfo {author} {\bibfnamefont {A.~P.}\
  \bibnamefont {Schnyder}}, \ and\ \bibinfo {author} {\bibfnamefont
  {P.}~\bibnamefont {Wahl}},\ }\href {\doibase 10.1038/ncomms7633} {\bibfield
  {journal} {\bibinfo  {journal} {Nat. Commun.}\ }\textbf {\bibinfo {volume}
  {6}},\ \bibinfo {pages} {1} (\bibinfo {year} {2015})}\BibitemShut {NoStop}%
\bibitem [{\citenamefont {Chang}\ \emph {et~al.}(2016)\citenamefont {Chang},
  \citenamefont {Chen}, \citenamefont {Bian}, \citenamefont {Huang},
  \citenamefont {Zheng}, \citenamefont {Neupert}, \citenamefont {Sankar},
  \citenamefont {Xu}, \citenamefont {Belopolski}, \citenamefont {Chang},
  \citenamefont {Wang}, \citenamefont {Chou}, \citenamefont {Bansil},
  \citenamefont {Jeng}, \citenamefont {Lin},\ and\ \citenamefont
  {Hasan}}]{Chang2016}%
  \BibitemOpen
  \bibfield  {author} {\bibinfo {author} {\bibfnamefont {T.-R.}\ \bibnamefont
  {Chang}}, \bibinfo {author} {\bibfnamefont {P.-J.}\ \bibnamefont {Chen}},
  \bibinfo {author} {\bibfnamefont {G.}~\bibnamefont {Bian}}, \bibinfo {author}
  {\bibfnamefont {S.-M.}\ \bibnamefont {Huang}}, \bibinfo {author}
  {\bibfnamefont {H.}~\bibnamefont {Zheng}}, \bibinfo {author} {\bibfnamefont
  {T.}~\bibnamefont {Neupert}}, \bibinfo {author} {\bibfnamefont
  {R.}~\bibnamefont {Sankar}}, \bibinfo {author} {\bibfnamefont {S.-Y.}\
  \bibnamefont {Xu}}, \bibinfo {author} {\bibfnamefont {I.}~\bibnamefont
  {Belopolski}}, \bibinfo {author} {\bibfnamefont {G.}~\bibnamefont {Chang}},
  \bibinfo {author} {\bibfnamefont {B.}~\bibnamefont {Wang}}, \bibinfo {author}
  {\bibfnamefont {F.}~\bibnamefont {Chou}}, \bibinfo {author} {\bibfnamefont
  {A.}~\bibnamefont {Bansil}}, \bibinfo {author} {\bibfnamefont {H.-T.}\
  \bibnamefont {Jeng}}, \bibinfo {author} {\bibfnamefont {H.}~\bibnamefont
  {Lin}}, \ and\ \bibinfo {author} {\bibfnamefont {M.~Z.}\ \bibnamefont
  {Hasan}},\ }\href {\doibase 10.1103/PhysRevB.93.245130} {\bibfield  {journal}
  {\bibinfo  {journal} {Phy. Rev. B}\ }\textbf {\bibinfo {volume} {93}},\
  \bibinfo {pages} {245130} (\bibinfo {year} {2016})}\BibitemShut {NoStop}%
\bibitem [{\citenamefont {Ka\ifmmode \check{c}\else
  \v{c}\fi{}mar\ifmmode~\check{c}\else \v{c}\fi{}\'{\i}k}\ \emph
  {et~al.}(2016)\citenamefont {Ka\ifmmode \check{c}\else
  \v{c}\fi{}mar\ifmmode~\check{c}\else \v{c}\fi{}\'{\i}k}, \citenamefont
  {Pribulov\'a}, \citenamefont {Samuely}, \citenamefont {Szab\'o},
  \citenamefont {Cambel}, \citenamefont {\ifmmode~\check{S}\else
  \v{S}\fi{}olt\'ys}, \citenamefont {Herrera}, \citenamefont {Suderow},
  \citenamefont {Correa-Orellana}, \citenamefont {Prabhakaran},\ and\
  \citenamefont {Samuely}}]{PhysRevB.93.144502}%
  \BibitemOpen
  \bibfield  {author} {\bibinfo {author} {\bibfnamefont {J.}~\bibnamefont
  {Ka\ifmmode \check{c}\else \v{c}\fi{}mar\ifmmode~\check{c}\else
  \v{c}\fi{}\'{\i}k}}, \bibinfo {author} {\bibfnamefont {Z.}~\bibnamefont
  {Pribulov\'a}}, \bibinfo {author} {\bibfnamefont {T.}~\bibnamefont
  {Samuely}}, \bibinfo {author} {\bibfnamefont {P.}~\bibnamefont {Szab\'o}},
  \bibinfo {author} {\bibfnamefont {V.}~\bibnamefont {Cambel}}, \bibinfo
  {author} {\bibfnamefont {J.}~\bibnamefont {\ifmmode~\check{S}\else
  \v{S}\fi{}olt\'ys}}, \bibinfo {author} {\bibfnamefont {E.}~\bibnamefont
  {Herrera}}, \bibinfo {author} {\bibfnamefont {H.}~\bibnamefont {Suderow}},
  \bibinfo {author} {\bibfnamefont {A.}~\bibnamefont {Correa-Orellana}},
  \bibinfo {author} {\bibfnamefont {D.}~\bibnamefont {Prabhakaran}}, \ and\
  \bibinfo {author} {\bibfnamefont {P.}~\bibnamefont {Samuely}},\ }\href
  {\doibase 10.1103/PhysRevB.93.144502} {\bibfield  {journal} {\bibinfo
  {journal} {Phys. Rev. B}\ }\textbf {\bibinfo {volume} {93}},\ \bibinfo
  {pages} {144502} (\bibinfo {year} {2016})}\BibitemShut {NoStop}%
\bibitem [{\citenamefont {Imai}\ \emph {et~al.}(2012)\citenamefont {Imai},
  \citenamefont {Nabeshima}, \citenamefont {Yoshinaka}, \citenamefont
  {Miyatani}, \citenamefont {Kondo}, \citenamefont {Komiya}, \citenamefont
  {Tsukada},\ and\ \citenamefont {Maeda}}]{Imai2012}%
  \BibitemOpen
  \bibfield  {author} {\bibinfo {author} {\bibfnamefont {Y.}~\bibnamefont
  {Imai}}, \bibinfo {author} {\bibfnamefont {F.}~\bibnamefont {Nabeshima}},
  \bibinfo {author} {\bibfnamefont {T.}~\bibnamefont {Yoshinaka}}, \bibinfo
  {author} {\bibfnamefont {K.}~\bibnamefont {Miyatani}}, \bibinfo {author}
  {\bibfnamefont {R.}~\bibnamefont {Kondo}}, \bibinfo {author} {\bibfnamefont
  {S.}~\bibnamefont {Komiya}}, \bibinfo {author} {\bibfnamefont
  {I.}~\bibnamefont {Tsukada}}, \ and\ \bibinfo {author} {\bibfnamefont
  {A.}~\bibnamefont {Maeda}},\ }\href {\doibase 10.1143/jpsj.81.113708}
  {\bibfield  {journal} {\bibinfo  {journal} {J. Phys. Soc. Jpn.}\ }\textbf
  {\bibinfo {volume} {81}},\ \bibinfo {pages} {113708} (\bibinfo {year}
  {2012})}\BibitemShut {NoStop}%
\bibitem [{\citenamefont {Margine}\ and\ \citenamefont
  {Giustino}(2013)}]{PhysRevB.87.024505}%
  \BibitemOpen
  \bibfield  {author} {\bibinfo {author} {\bibfnamefont {E.~R.}\ \bibnamefont
  {Margine}}\ and\ \bibinfo {author} {\bibfnamefont {F.}~\bibnamefont
  {Giustino}},\ }\href {\doibase 10.1103/PhysRevB.87.024505} {\bibfield
  {journal} {\bibinfo  {journal} {Phys. Rev. B}\ }\textbf {\bibinfo {volume}
  {87}},\ \bibinfo {pages} {024505} (\bibinfo {year} {2013})}\BibitemShut
  {NoStop}%
\bibitem [{\citenamefont {Zheng}\ and\ \citenamefont
  {Margine}(2017)}]{PhysRevB.95.014512}%
  \BibitemOpen
  \bibfield  {author} {\bibinfo {author} {\bibfnamefont {J.-J.}\ \bibnamefont
  {Zheng}}\ and\ \bibinfo {author} {\bibfnamefont {E.~R.}\ \bibnamefont
  {Margine}},\ }\href {\doibase 10.1103/PhysRevB.95.014512} {\bibfield
  {journal} {\bibinfo  {journal} {Phys. Rev. B}\ }\textbf {\bibinfo {volume}
  {95}},\ \bibinfo {pages} {014512} (\bibinfo {year} {2017})}\BibitemShut
  {NoStop}%
\bibitem [{\citenamefont {Zhang}\ \emph {et~al.}(2019)\citenamefont {Zhang},
  \citenamefont {Zhao},\ and\ \citenamefont {Liu}}]{PhysRevB.100.104527}%
  \BibitemOpen
  \bibfield  {author} {\bibinfo {author} {\bibfnamefont {X.}~\bibnamefont
  {Zhang}}, \bibinfo {author} {\bibfnamefont {M.}~\bibnamefont {Zhao}}, \ and\
  \bibinfo {author} {\bibfnamefont {F.}~\bibnamefont {Liu}},\ }\href {\doibase
  10.1103/PhysRevB.100.104527} {\bibfield  {journal} {\bibinfo  {journal}
  {Phys. Rev. B}\ }\textbf {\bibinfo {volume} {100}},\ \bibinfo {pages}
  {104527} (\bibinfo {year} {2019})}\BibitemShut {NoStop}%
\bibitem [{\citenamefont {Wang}\ and\ \citenamefont
  {Margine}(2017)}]{Wang2017}%
  \BibitemOpen
  \bibfield  {author} {\bibinfo {author} {\bibfnamefont {B.~T.}\ \bibnamefont
  {Wang}}\ and\ \bibinfo {author} {\bibfnamefont {E.~R.}\ \bibnamefont
  {Margine}},\ }\href {\doibase 10.1088/1361-648X/aa7a4b} {\bibfield  {journal}
  {\bibinfo  {journal} {J. Phys.: Condens. Matter.}\ }\textbf {\bibinfo
  {volume} {29}},\ \bibinfo {pages} {325501} (\bibinfo {year}
  {2017})}\BibitemShut {NoStop}%
\bibitem [{\citenamefont {Li}\ and\ \citenamefont {Xu}(2019)}]{Li2019}%
  \BibitemOpen
  \bibfield  {author} {\bibinfo {author} {\bibfnamefont {Y.}~\bibnamefont
  {Li}}\ and\ \bibinfo {author} {\bibfnamefont {Z.-A.}\ \bibnamefont {Xu}},\
  }\href {\doibase 10.1002/qute.201800112} {\bibfield  {journal} {\bibinfo
  {journal} {Adv. Quantum Technol.}\ }\textbf {\bibinfo {volume} {2}},\
  \bibinfo {pages} {1800112} (\bibinfo {year} {2019})}\BibitemShut {NoStop}%
\bibitem [{\citenamefont {Xu}\ \emph {et~al.}(2019)\citenamefont {Xu},
  \citenamefont {Wang}, \citenamefont {Wang}, \citenamefont {Jiang},
  \citenamefont {Shen}, \citenamefont {Gao}, \citenamefont {Ye},\ and\
  \citenamefont {Qiao}}]{Xu2019}%
  \BibitemOpen
  \bibfield  {author} {\bibinfo {author} {\bibfnamefont {T.}~\bibnamefont
  {Xu}}, \bibinfo {author} {\bibfnamefont {B.~T.}\ \bibnamefont {Wang}},
  \bibinfo {author} {\bibfnamefont {M.}~\bibnamefont {Wang}}, \bibinfo {author}
  {\bibfnamefont {Q.}~\bibnamefont {Jiang}}, \bibinfo {author} {\bibfnamefont
  {X.~P.}\ \bibnamefont {Shen}}, \bibinfo {author} {\bibfnamefont
  {B.}~\bibnamefont {Gao}}, \bibinfo {author} {\bibfnamefont {M.}~\bibnamefont
  {Ye}}, \ and\ \bibinfo {author} {\bibfnamefont {S.}~\bibnamefont {Qiao}},\
  }\href {\doibase 10.1103/PhysRevB.100.161109} {\bibfield  {journal} {\bibinfo
   {journal} {Phys. Rev. B}\ }\textbf {\bibinfo {volume} {100}},\ \bibinfo
  {pages} {161109} (\bibinfo {year} {2019})}\BibitemShut {NoStop}%
\bibitem [{\citenamefont {Guan}\ \emph {et~al.}(2019)\citenamefont {Guan},
  \citenamefont {Kong}, \citenamefont {Zhou}, \citenamefont {Zhong},
  \citenamefont {Li}, \citenamefont {Liu}, \citenamefont {Tang}, \citenamefont
  {Yan}, \citenamefont {Yang}, \citenamefont {Huang}, \citenamefont {Shi},
  \citenamefont {Qian}, \citenamefont {Weng}, \citenamefont {Sun},\ and\
  \citenamefont {Ding}}]{GUAN20191215}%
  \BibitemOpen
  \bibfield  {author} {\bibinfo {author} {\bibfnamefont {J.-Y.}\ \bibnamefont
  {Guan}}, \bibinfo {author} {\bibfnamefont {L.}~\bibnamefont {Kong}}, \bibinfo
  {author} {\bibfnamefont {L.-Q.}\ \bibnamefont {Zhou}}, \bibinfo {author}
  {\bibfnamefont {Y.-G.}\ \bibnamefont {Zhong}}, \bibinfo {author}
  {\bibfnamefont {H.}~\bibnamefont {Li}}, \bibinfo {author} {\bibfnamefont
  {H.-J.}\ \bibnamefont {Liu}}, \bibinfo {author} {\bibfnamefont {C.-Y.}\
  \bibnamefont {Tang}}, \bibinfo {author} {\bibfnamefont {D.-Y.}\ \bibnamefont
  {Yan}}, \bibinfo {author} {\bibfnamefont {F.-Z.}\ \bibnamefont {Yang}},
  \bibinfo {author} {\bibfnamefont {Y.-B.}\ \bibnamefont {Huang}}, \bibinfo
  {author} {\bibfnamefont {Y.-G.}\ \bibnamefont {Shi}}, \bibinfo {author}
  {\bibfnamefont {T.}~\bibnamefont {Qian}}, \bibinfo {author} {\bibfnamefont
  {H.-M.}\ \bibnamefont {Weng}}, \bibinfo {author} {\bibfnamefont {Y.-J.}\
  \bibnamefont {Sun}}, \ and\ \bibinfo {author} {\bibfnamefont
  {H.}~\bibnamefont {Ding}},\ }\href {\doibase
  https://doi.org/10.1016/j.scib.2019.07.019} {\bibfield  {journal} {\bibinfo
  {journal} {Sci. Bull.}\ }\textbf {\bibinfo {volume} {64}},\ \bibinfo {pages}
  {1215 } (\bibinfo {year} {2019})}\BibitemShut {NoStop}%
\bibitem [{\citenamefont {Prist\'a\ifmmode~\check{s}\else \v{s}\fi{}}\ \emph
  {et~al.}(2018)\citenamefont {Prist\'a\ifmmode~\check{s}\else \v{s}\fi{}},
  \citenamefont {Orend\'a\ifmmode~\check{c}\else \v{c}\fi{}}, \citenamefont
  {Gab\'ani}, \citenamefont {Ka\ifmmode \check{c}\else
  \v{c}\fi{}mar\ifmmode~\check{c}\else \v{c}\fi{}\'{\i}k}, \citenamefont
  {Ga\ifmmode~\check{z}\else \v{z}\fi{}o}, \citenamefont {Pribulov\'a},
  \citenamefont {Correa-Orellana}, \citenamefont {Herrera}, \citenamefont
  {Suderow},\ and\ \citenamefont {Samuely}}]{PhysRevB.97.134505}%
  \BibitemOpen
  \bibfield  {author} {\bibinfo {author} {\bibfnamefont {G.}~\bibnamefont
  {Prist\'a\ifmmode~\check{s}\else \v{s}\fi{}}}, \bibinfo {author}
  {\bibfnamefont {M.}~\bibnamefont {Orend\'a\ifmmode~\check{c}\else
  \v{c}\fi{}}}, \bibinfo {author} {\bibfnamefont {S.}~\bibnamefont {Gab\'ani}},
  \bibinfo {author} {\bibfnamefont {J.}~\bibnamefont {Ka\ifmmode \check{c}\else
  \v{c}\fi{}mar\ifmmode~\check{c}\else \v{c}\fi{}\'{\i}k}}, \bibinfo {author}
  {\bibfnamefont {E.}~\bibnamefont {Ga\ifmmode~\check{z}\else \v{z}\fi{}o}},
  \bibinfo {author} {\bibfnamefont {Z.}~\bibnamefont {Pribulov\'a}}, \bibinfo
  {author} {\bibfnamefont {A.}~\bibnamefont {Correa-Orellana}}, \bibinfo
  {author} {\bibfnamefont {E.}~\bibnamefont {Herrera}}, \bibinfo {author}
  {\bibfnamefont {H.}~\bibnamefont {Suderow}}, \ and\ \bibinfo {author}
  {\bibfnamefont {P.}~\bibnamefont {Samuely}},\ }\href {\doibase
  10.1103/PhysRevB.97.134505} {\bibfield  {journal} {\bibinfo  {journal} {Phys.
  Rev. B}\ }\textbf {\bibinfo {volume} {97}},\ \bibinfo {pages} {134505}
  (\bibinfo {year} {2018})}\BibitemShut {NoStop}%
\bibitem [{\citenamefont {Denisov}\ \emph {et~al.}(2017)\citenamefont
  {Denisov}, \citenamefont {Matetskiy}, \citenamefont {Tupkalo}, \citenamefont
  {Zotov},\ and\ \citenamefont {Saranin}}]{Denisov2017}%
  \BibitemOpen
  \bibfield  {author} {\bibinfo {author} {\bibfnamefont {N.}~\bibnamefont
  {Denisov}}, \bibinfo {author} {\bibfnamefont {A.}~\bibnamefont {Matetskiy}},
  \bibinfo {author} {\bibfnamefont {A.}~\bibnamefont {Tupkalo}}, \bibinfo
  {author} {\bibfnamefont {A.}~\bibnamefont {Zotov}}, \ and\ \bibinfo {author}
  {\bibfnamefont {A.}~\bibnamefont {Saranin}},\ }\href {\doibase
  10.1016/j.apsusc.2016.12.239} {\bibfield  {journal} {\bibinfo  {journal}
  {Appl. Surf. Sci.}\ }\textbf {\bibinfo {volume} {401}},\ \bibinfo {pages}
  {142} (\bibinfo {year} {2017})}\BibitemShut {NoStop}%
\bibitem [{\citenamefont {Troullier}\ and\ \citenamefont
  {Martins}(1991)}]{PhysRevB.43.1993}%
  \BibitemOpen
  \bibfield  {author} {\bibinfo {author} {\bibfnamefont {N.}~\bibnamefont
  {Troullier}}\ and\ \bibinfo {author} {\bibfnamefont {J.~L.}\ \bibnamefont
  {Martins}},\ }\href {\doibase 10.1103/physrevb.43.1993} {\bibfield  {journal}
  {\bibinfo  {journal} {Phys. Rev. B}\ }\textbf {\bibinfo {volume} {43}},\
  \bibinfo {pages} {1993} (\bibinfo {year} {1991})}\BibitemShut {NoStop}%
\bibitem [{\citenamefont {Fuchs}\ and\ \citenamefont
  {Scheffler}(1999)}]{FUCHS199967}%
  \BibitemOpen
  \bibfield  {author} {\bibinfo {author} {\bibfnamefont {M.}~\bibnamefont
  {Fuchs}}\ and\ \bibinfo {author} {\bibfnamefont {M.}~\bibnamefont
  {Scheffler}},\ }\href {\doibase 10.1016/S0010-4655(98)00201-X} {\bibfield
  {journal} {\bibinfo  {journal} {Comput. Phys. Commun.}\ }\textbf {\bibinfo
  {volume} {119}},\ \bibinfo {pages} {67 } (\bibinfo {year}
  {1999})}\BibitemShut {NoStop}%
\bibitem [{\citenamefont {Giannozzi}\ \emph {et~al.}(2009)\citenamefont
  {Giannozzi}, \citenamefont {Baroni}, \citenamefont {Bonini}, \citenamefont
  {Calandra}, \citenamefont {Car}, \citenamefont {Cavazzoni}, \citenamefont
  {Ceresoli}, \citenamefont {Chiarotti}, \citenamefont {Cococcioni},
  \citenamefont {Dabo}, \citenamefont {Corso}, \citenamefont {de~Gironcoli},
  \citenamefont {Fabris}, \citenamefont {Fratesi}, \citenamefont {Gebauer},
  \citenamefont {Gerstmann}, \citenamefont {Gougoussis}, \citenamefont
  {Kokalj}, \citenamefont {Lazzeri}, \citenamefont {Martin-Samos},
  \citenamefont {Marzari}, \citenamefont {Mauri}, \citenamefont {Mazzarello},
  \citenamefont {Paolini}, \citenamefont {Pasquarello}, \citenamefont
  {Paulatto}, \citenamefont {Sbraccia}, \citenamefont {Scandolo}, \citenamefont
  {Sclauzero}, \citenamefont {Seitsonen}, \citenamefont {Smogunov},
  \citenamefont {Umari},\ and\ \citenamefont {Wentzcovitch}}]{qe2009}%
  \BibitemOpen
  \bibfield  {author} {\bibinfo {author} {\bibfnamefont {P.}~\bibnamefont
  {Giannozzi}}, \bibinfo {author} {\bibfnamefont {S.}~\bibnamefont {Baroni}},
  \bibinfo {author} {\bibfnamefont {N.}~\bibnamefont {Bonini}}, \bibinfo
  {author} {\bibfnamefont {M.}~\bibnamefont {Calandra}}, \bibinfo {author}
  {\bibfnamefont {R.}~\bibnamefont {Car}}, \bibinfo {author} {\bibfnamefont
  {C.}~\bibnamefont {Cavazzoni}}, \bibinfo {author} {\bibfnamefont
  {D.}~\bibnamefont {Ceresoli}}, \bibinfo {author} {\bibfnamefont {G.~L.}\
  \bibnamefont {Chiarotti}}, \bibinfo {author} {\bibfnamefont {M.}~\bibnamefont
  {Cococcioni}}, \bibinfo {author} {\bibfnamefont {I.}~\bibnamefont {Dabo}},
  \bibinfo {author} {\bibfnamefont {A.~D.}\ \bibnamefont {Corso}}, \bibinfo
  {author} {\bibfnamefont {S.}~\bibnamefont {de~Gironcoli}}, \bibinfo {author}
  {\bibfnamefont {S.}~\bibnamefont {Fabris}}, \bibinfo {author} {\bibfnamefont
  {G.}~\bibnamefont {Fratesi}}, \bibinfo {author} {\bibfnamefont
  {R.}~\bibnamefont {Gebauer}}, \bibinfo {author} {\bibfnamefont
  {U.}~\bibnamefont {Gerstmann}}, \bibinfo {author} {\bibfnamefont
  {C.}~\bibnamefont {Gougoussis}}, \bibinfo {author} {\bibfnamefont
  {A.}~\bibnamefont {Kokalj}}, \bibinfo {author} {\bibfnamefont
  {M.}~\bibnamefont {Lazzeri}}, \bibinfo {author} {\bibfnamefont
  {L.}~\bibnamefont {Martin-Samos}}, \bibinfo {author} {\bibfnamefont
  {N.}~\bibnamefont {Marzari}}, \bibinfo {author} {\bibfnamefont
  {F.}~\bibnamefont {Mauri}}, \bibinfo {author} {\bibfnamefont
  {R.}~\bibnamefont {Mazzarello}}, \bibinfo {author} {\bibfnamefont
  {S.}~\bibnamefont {Paolini}}, \bibinfo {author} {\bibfnamefont
  {A.}~\bibnamefont {Pasquarello}}, \bibinfo {author} {\bibfnamefont
  {L.}~\bibnamefont {Paulatto}}, \bibinfo {author} {\bibfnamefont
  {C.}~\bibnamefont {Sbraccia}}, \bibinfo {author} {\bibfnamefont
  {S.}~\bibnamefont {Scandolo}}, \bibinfo {author} {\bibfnamefont
  {G.}~\bibnamefont {Sclauzero}}, \bibinfo {author} {\bibfnamefont {A.~P.}\
  \bibnamefont {Seitsonen}}, \bibinfo {author} {\bibfnamefont {A.}~\bibnamefont
  {Smogunov}}, \bibinfo {author} {\bibfnamefont {P.}~\bibnamefont {Umari}}, \
  and\ \bibinfo {author} {\bibfnamefont {R.~M.}\ \bibnamefont {Wentzcovitch}},\
  }\href {\doibase 10.1088/0953-8984/21/39/395502} {\bibfield  {journal}
  {\bibinfo  {journal} {J. Phys.: Condens. Matter.}\ }\textbf {\bibinfo
  {volume} {21}},\ \bibinfo {pages} {395502} (\bibinfo {year}
  {2009})}\BibitemShut {NoStop}%
\bibitem [{\citenamefont {Giannozzi}\ \emph {et~al.}(2017)\citenamefont
  {Giannozzi}, \citenamefont {Andreussi}, \citenamefont {Brumme}, \citenamefont
  {Bunau}, \citenamefont {Nardelli}, \citenamefont {Calandra}, \citenamefont
  {Car}, \citenamefont {Cavazzoni}, \citenamefont {Ceresoli}, \citenamefont
  {Cococcioni}, \citenamefont {Colonna}, \citenamefont {Carnimeo},
  \citenamefont {Corso}, \citenamefont {de~Gironcoli}, \citenamefont {Delugas},
  \citenamefont {Jr}, \citenamefont {Ferretti}, \citenamefont {Floris},
  \citenamefont {Fratesi}, \citenamefont {Fugallo}, \citenamefont {Gebauer},
  \citenamefont {Gerstmann}, \citenamefont {Giustino}, \citenamefont {Gorni},
  \citenamefont {Jia}, \citenamefont {Kawamura}, \citenamefont {Ko},
  \citenamefont {Kokalj}, \citenamefont {K¨¹c¨¹kbenli}, \citenamefont
  {Lazzeri}, \citenamefont {Marsili}, \citenamefont {Marzari}, \citenamefont
  {Mauri}, \citenamefont {Nguyen}, \citenamefont {Nguyen}, \citenamefont {de-la
  Roza}, \citenamefont {Paulatto}, \citenamefont {Ponc¨¦}, \citenamefont
  {Rocca}, \citenamefont {Sabatini}, \citenamefont {Santra}, \citenamefont
  {Schlipf}, \citenamefont {Seitsonen}, \citenamefont {Smogunov}, \citenamefont
  {Timrov}, \citenamefont {Thonhauser}, \citenamefont {Umari}, \citenamefont
  {Vast}, \citenamefont {Wu},\ and\ \citenamefont {Baroni}}]{qe2017}%
  \BibitemOpen
  \bibfield  {author} {\bibinfo {author} {\bibfnamefont {P.}~\bibnamefont
  {Giannozzi}}, \bibinfo {author} {\bibfnamefont {O.}~\bibnamefont
  {Andreussi}}, \bibinfo {author} {\bibfnamefont {T.}~\bibnamefont {Brumme}},
  \bibinfo {author} {\bibfnamefont {O.}~\bibnamefont {Bunau}}, \bibinfo
  {author} {\bibfnamefont {M.~B.}\ \bibnamefont {Nardelli}}, \bibinfo {author}
  {\bibfnamefont {M.}~\bibnamefont {Calandra}}, \bibinfo {author}
  {\bibfnamefont {R.}~\bibnamefont {Car}}, \bibinfo {author} {\bibfnamefont
  {C.}~\bibnamefont {Cavazzoni}}, \bibinfo {author} {\bibfnamefont
  {D.}~\bibnamefont {Ceresoli}}, \bibinfo {author} {\bibfnamefont
  {M.}~\bibnamefont {Cococcioni}}, \bibinfo {author} {\bibfnamefont
  {N.}~\bibnamefont {Colonna}}, \bibinfo {author} {\bibfnamefont
  {I.}~\bibnamefont {Carnimeo}}, \bibinfo {author} {\bibfnamefont {A.~D.}\
  \bibnamefont {Corso}}, \bibinfo {author} {\bibfnamefont {S.}~\bibnamefont
  {de~Gironcoli}}, \bibinfo {author} {\bibfnamefont {P.}~\bibnamefont
  {Delugas}}, \bibinfo {author} {\bibfnamefont {R.~A.~D.}\ \bibnamefont {Jr}},
  \bibinfo {author} {\bibfnamefont {A.}~\bibnamefont {Ferretti}}, \bibinfo
  {author} {\bibfnamefont {A.}~\bibnamefont {Floris}}, \bibinfo {author}
  {\bibfnamefont {G.}~\bibnamefont {Fratesi}}, \bibinfo {author} {\bibfnamefont
  {G.}~\bibnamefont {Fugallo}}, \bibinfo {author} {\bibfnamefont
  {R.}~\bibnamefont {Gebauer}}, \bibinfo {author} {\bibfnamefont
  {U.}~\bibnamefont {Gerstmann}}, \bibinfo {author} {\bibfnamefont
  {F.}~\bibnamefont {Giustino}}, \bibinfo {author} {\bibfnamefont
  {T.}~\bibnamefont {Gorni}}, \bibinfo {author} {\bibfnamefont
  {J.}~\bibnamefont {Jia}}, \bibinfo {author} {\bibfnamefont {M.}~\bibnamefont
  {Kawamura}}, \bibinfo {author} {\bibfnamefont {H.-Y.}\ \bibnamefont {Ko}},
  \bibinfo {author} {\bibfnamefont {A.}~\bibnamefont {Kokalj}}, \bibinfo
  {author} {\bibfnamefont {E.}~\bibnamefont {K¨¹c¨¹kbenli}}, \bibinfo {author}
  {\bibfnamefont {M.}~\bibnamefont {Lazzeri}}, \bibinfo {author} {\bibfnamefont
  {M.}~\bibnamefont {Marsili}}, \bibinfo {author} {\bibfnamefont
  {N.}~\bibnamefont {Marzari}}, \bibinfo {author} {\bibfnamefont
  {F.}~\bibnamefont {Mauri}}, \bibinfo {author} {\bibfnamefont {N.~L.}\
  \bibnamefont {Nguyen}}, \bibinfo {author} {\bibfnamefont {H.-V.}\
  \bibnamefont {Nguyen}}, \bibinfo {author} {\bibfnamefont {A.~O.}\
  \bibnamefont {de-la Roza}}, \bibinfo {author} {\bibfnamefont
  {L.}~\bibnamefont {Paulatto}}, \bibinfo {author} {\bibfnamefont
  {S.}~\bibnamefont {Ponc¨¦}}, \bibinfo {author} {\bibfnamefont
  {D.}~\bibnamefont {Rocca}}, \bibinfo {author} {\bibfnamefont
  {R.}~\bibnamefont {Sabatini}}, \bibinfo {author} {\bibfnamefont
  {B.}~\bibnamefont {Santra}}, \bibinfo {author} {\bibfnamefont
  {M.}~\bibnamefont {Schlipf}}, \bibinfo {author} {\bibfnamefont {A.~P.}\
  \bibnamefont {Seitsonen}}, \bibinfo {author} {\bibfnamefont {A.}~\bibnamefont
  {Smogunov}}, \bibinfo {author} {\bibfnamefont {I.}~\bibnamefont {Timrov}},
  \bibinfo {author} {\bibfnamefont {T.}~\bibnamefont {Thonhauser}}, \bibinfo
  {author} {\bibfnamefont {P.}~\bibnamefont {Umari}}, \bibinfo {author}
  {\bibfnamefont {N.}~\bibnamefont {Vast}}, \bibinfo {author} {\bibfnamefont
  {X.}~\bibnamefont {Wu}}, \ and\ \bibinfo {author} {\bibfnamefont
  {S.}~\bibnamefont {Baroni}},\ }\href {\doibase 10.1088/1361-648X/aa8f79}
  {\bibfield  {journal} {\bibinfo  {journal} {J. Phys.: Condens. Matter.}\
  }\textbf {\bibinfo {volume} {29}},\ \bibinfo {pages} {465901} (\bibinfo
  {year} {2017})}\BibitemShut {NoStop}%
\bibitem [{\citenamefont {Perdew}\ \emph {et~al.}(1996)\citenamefont {Perdew},
  \citenamefont {Burke},\ and\ \citenamefont
  {Ernzerhof}}]{PhysRevLett.77.3865}%
  \BibitemOpen
  \bibfield  {author} {\bibinfo {author} {\bibfnamefont {J.~P.}\ \bibnamefont
  {Perdew}}, \bibinfo {author} {\bibfnamefont {K.}~\bibnamefont {Burke}}, \
  and\ \bibinfo {author} {\bibfnamefont {M.}~\bibnamefont {Ernzerhof}},\ }\href
  {\doibase 10.1103/PhysRevLett.77.3865} {\bibfield  {journal} {\bibinfo
  {journal} {Phys. Rev. Lett.}\ }\textbf {\bibinfo {volume} {77}},\ \bibinfo
  {pages} {3865} (\bibinfo {year} {1996})}\BibitemShut {NoStop}%
\bibitem [{\citenamefont {Baroni}\ \emph {et~al.}(2001)\citenamefont {Baroni},
  \citenamefont {De~Gironcoli}, \citenamefont {Dal~Corso},\ and\ \citenamefont
  {Giannozzi}}]{baroni2001phonons}%
  \BibitemOpen
  \bibfield  {author} {\bibinfo {author} {\bibfnamefont {S.}~\bibnamefont
  {Baroni}}, \bibinfo {author} {\bibfnamefont {S.}~\bibnamefont
  {De~Gironcoli}}, \bibinfo {author} {\bibfnamefont {A.}~\bibnamefont
  {Dal~Corso}}, \ and\ \bibinfo {author} {\bibfnamefont {P.}~\bibnamefont
  {Giannozzi}},\ }\href {\doibase 10.1103/RevModPhys.73.515} {\bibfield
  {journal} {\bibinfo  {journal} {Rev. Mod. Phys.}\ }\textbf {\bibinfo {volume}
  {73}},\ \bibinfo {pages} {515} (\bibinfo {year} {2001})}\BibitemShut
  {NoStop}%
\bibitem [{\citenamefont {Grimvall}(1981)}]{grimvall1981electron}%
  \BibitemOpen
  \bibfield  {author} {\bibinfo {author} {\bibfnamefont {G.}~\bibnamefont
  {Grimvall}},\ }\href {\doibase 10.1103/PhysRev.99.1140} {\emph {\bibinfo
  {title} {The electron-phonon interaction in metals}}},\ Vol.~\bibinfo
  {volume} {8}\ (\bibinfo  {publisher} {North-Holland Amsterdam},\ \bibinfo
  {year} {1981})\BibitemShut {NoStop}%
\bibitem [{\citenamefont {Giustino}(2017)}]{giustino2017electron}%
  \BibitemOpen
  \bibfield  {author} {\bibinfo {author} {\bibfnamefont {F.}~\bibnamefont
  {Giustino}},\ }\href {\doibase 10.1103/RevModPhys.89.015003} {\bibfield
  {journal} {\bibinfo  {journal} {Rev. Mod. Phys.}\ }\textbf {\bibinfo {volume}
  {89}},\ \bibinfo {pages} {015003} (\bibinfo {year} {2017})}\BibitemShut
  {NoStop}%
\bibitem [{\citenamefont {Allen}\ and\ \citenamefont
  {Dynes}(1975{\natexlab{a}})}]{allen1975transition}%
  \BibitemOpen
  \bibfield  {author} {\bibinfo {author} {\bibfnamefont {P.~B.}\ \bibnamefont
  {Allen}}\ and\ \bibinfo {author} {\bibfnamefont {R.}~\bibnamefont {Dynes}},\
  }\href {\doibase 10.1103/PhysRevB.12.905} {\bibfield  {journal} {\bibinfo
  {journal} {Phys. Rev. B}\ }\textbf {\bibinfo {volume} {12}},\ \bibinfo
  {pages} {905} (\bibinfo {year} {1975}{\natexlab{a}})}\BibitemShut {NoStop}%
\bibitem [{\citenamefont {Eliashberg}(1960)}]{osti_7354388}%
  \BibitemOpen
  \bibfield  {author} {\bibinfo {author} {\bibfnamefont {G.}~\bibnamefont
  {Eliashberg}},\ }\href {http://www.jetp.ac.ru/cgi-bin/dn/e_011_03_0696.pdf}
  {\bibfield  {journal} {\bibinfo  {journal} {Sov. Phys. JETP}\ }\textbf
  {\bibinfo {volume} {11}},\ \bibinfo {pages} {696} (\bibinfo {year}
  {1960})}\BibitemShut {NoStop}%
\bibitem [{\citenamefont {Allen}\ and\ \citenamefont
  {Dynes}(1975{\natexlab{b}})}]{PhysRevB.12.905}%
  \BibitemOpen
  \bibfield  {author} {\bibinfo {author} {\bibfnamefont {P.~B.}\ \bibnamefont
  {Allen}}\ and\ \bibinfo {author} {\bibfnamefont {R.~C.}\ \bibnamefont
  {Dynes}},\ }\href {\doibase 10.1103/PhysRevB.12.905} {\bibfield  {journal}
  {\bibinfo  {journal} {Phys. Rev. B}\ }\textbf {\bibinfo {volume} {12}},\
  \bibinfo {pages} {905} (\bibinfo {year} {1975}{\natexlab{b}})}\BibitemShut
  {NoStop}%
\bibitem [{\citenamefont {Marzari}\ \emph {et~al.}(2012)\citenamefont
  {Marzari}, \citenamefont {Mostofi}, \citenamefont {Yates}, \citenamefont
  {Souza},\ and\ \citenamefont {Vanderbilt}}]{RevModPhys.84.1419}%
  \BibitemOpen
  \bibfield  {author} {\bibinfo {author} {\bibfnamefont {N.}~\bibnamefont
  {Marzari}}, \bibinfo {author} {\bibfnamefont {A.~A.}\ \bibnamefont
  {Mostofi}}, \bibinfo {author} {\bibfnamefont {J.~R.}\ \bibnamefont {Yates}},
  \bibinfo {author} {\bibfnamefont {I.}~\bibnamefont {Souza}}, \ and\ \bibinfo
  {author} {\bibfnamefont {D.}~\bibnamefont {Vanderbilt}},\ }\href {\doibase
  10.1103/RevModPhys.84.1419} {\bibfield  {journal} {\bibinfo  {journal} {Rev.
  Mod. Phys.}\ }\textbf {\bibinfo {volume} {84}},\ \bibinfo {pages} {1419}
  (\bibinfo {year} {2012})}\BibitemShut {NoStop}%
\bibitem [{\citenamefont {Mostofi}\ \emph {et~al.}(2014)\citenamefont
  {Mostofi}, \citenamefont {Yates}, \citenamefont {Pizzi}, \citenamefont {Lee},
  \citenamefont {Souza}, \citenamefont {Vanderbilt},\ and\ \citenamefont
  {Marzari}}]{MOSTOFI20142309}%
  \BibitemOpen
  \bibfield  {author} {\bibinfo {author} {\bibfnamefont {A.~A.}\ \bibnamefont
  {Mostofi}}, \bibinfo {author} {\bibfnamefont {J.~R.}\ \bibnamefont {Yates}},
  \bibinfo {author} {\bibfnamefont {G.}~\bibnamefont {Pizzi}}, \bibinfo
  {author} {\bibfnamefont {Y.-S.}\ \bibnamefont {Lee}}, \bibinfo {author}
  {\bibfnamefont {I.}~\bibnamefont {Souza}}, \bibinfo {author} {\bibfnamefont
  {D.}~\bibnamefont {Vanderbilt}}, \ and\ \bibinfo {author} {\bibfnamefont
  {N.}~\bibnamefont {Marzari}},\ }\href {\doibase 10.1016/j.cpc.2014.05.003}
  {\bibfield  {journal} {\bibinfo  {journal} {Comput. Phys. Commun.}\ }\textbf
  {\bibinfo {volume} {185}},\ \bibinfo {pages} {2309 } (\bibinfo {year}
  {2014})}\BibitemShut {NoStop}%
\bibitem [{\citenamefont {Pizzi}\ \emph {et~al.}(2020)\citenamefont {Pizzi},
  \citenamefont {Vitale}, \citenamefont {Arita}, \citenamefont {Bl¨¹gel},
  \citenamefont {Freimuth}, \citenamefont {G{\'{e}}ranton}, \citenamefont
  {Gibertini}, \citenamefont {Gresch}, \citenamefont {Johnson}, \citenamefont
  {Koretsune}, \citenamefont {Iba{\~{n}}ez-Azpiroz}, \citenamefont {Lee},
  \citenamefont {Lihm}, \citenamefont {Marchand}, \citenamefont {Marrazzo},
  \citenamefont {Mokrousov}, \citenamefont {Mustafa}, \citenamefont {Nohara},
  \citenamefont {Nomura}, \citenamefont {Paulatto}, \citenamefont
  {Ponc{\'{e}}}, \citenamefont {Ponweiser}, \citenamefont {Qiao}, \citenamefont
  {Th$\ddot{o}$le}, \citenamefont {Tsirkin}, \citenamefont {Wierzbowska},
  \citenamefont {Marzari}, \citenamefont {Vanderbilt}, \citenamefont {Souza},
  \citenamefont {Mostofi},\ and\ \citenamefont {Yates}}]{Pizzi_2020}%
  \BibitemOpen
  \bibfield  {author} {\bibinfo {author} {\bibfnamefont {G.}~\bibnamefont
  {Pizzi}}, \bibinfo {author} {\bibfnamefont {V.}~\bibnamefont {Vitale}},
  \bibinfo {author} {\bibfnamefont {R.}~\bibnamefont {Arita}}, \bibinfo
  {author} {\bibfnamefont {S.}~\bibnamefont {Bl¨¹gel}}, \bibinfo {author}
  {\bibfnamefont {F.}~\bibnamefont {Freimuth}}, \bibinfo {author}
  {\bibfnamefont {G.}~\bibnamefont {G{\'{e}}ranton}}, \bibinfo {author}
  {\bibfnamefont {M.}~\bibnamefont {Gibertini}}, \bibinfo {author}
  {\bibfnamefont {D.}~\bibnamefont {Gresch}}, \bibinfo {author} {\bibfnamefont
  {C.}~\bibnamefont {Johnson}}, \bibinfo {author} {\bibfnamefont
  {T.}~\bibnamefont {Koretsune}}, \bibinfo {author} {\bibfnamefont
  {J.}~\bibnamefont {Iba{\~{n}}ez-Azpiroz}}, \bibinfo {author} {\bibfnamefont
  {H.}~\bibnamefont {Lee}}, \bibinfo {author} {\bibfnamefont {J.-M.}\
  \bibnamefont {Lihm}}, \bibinfo {author} {\bibfnamefont {D.}~\bibnamefont
  {Marchand}}, \bibinfo {author} {\bibfnamefont {A.}~\bibnamefont {Marrazzo}},
  \bibinfo {author} {\bibfnamefont {Y.}~\bibnamefont {Mokrousov}}, \bibinfo
  {author} {\bibfnamefont {J.~I.}\ \bibnamefont {Mustafa}}, \bibinfo {author}
  {\bibfnamefont {Y.}~\bibnamefont {Nohara}}, \bibinfo {author} {\bibfnamefont
  {Y.}~\bibnamefont {Nomura}}, \bibinfo {author} {\bibfnamefont
  {L.}~\bibnamefont {Paulatto}}, \bibinfo {author} {\bibfnamefont
  {S.}~\bibnamefont {Ponc{\'{e}}}}, \bibinfo {author} {\bibfnamefont
  {T.}~\bibnamefont {Ponweiser}}, \bibinfo {author} {\bibfnamefont
  {J.}~\bibnamefont {Qiao}}, \bibinfo {author} {\bibfnamefont {F.}~\bibnamefont
  {Th$\ddot{o}$le}}, \bibinfo {author} {\bibfnamefont {S.~S.}\ \bibnamefont
  {Tsirkin}}, \bibinfo {author} {\bibfnamefont {M.}~\bibnamefont
  {Wierzbowska}}, \bibinfo {author} {\bibfnamefont {N.}~\bibnamefont
  {Marzari}}, \bibinfo {author} {\bibfnamefont {D.}~\bibnamefont {Vanderbilt}},
  \bibinfo {author} {\bibfnamefont {I.}~\bibnamefont {Souza}}, \bibinfo
  {author} {\bibfnamefont {A.~A.}\ \bibnamefont {Mostofi}}, \ and\ \bibinfo
  {author} {\bibfnamefont {J.~R.}\ \bibnamefont {Yates}},\ }\href {\doibase
  10.1088/1361-648X/ab51ff} {\bibfield  {journal} {\bibinfo  {journal} {J.
  Phys.: Condens. Matter.}\ }\textbf {\bibinfo {volume} {32}},\ \bibinfo
  {pages} {165902} (\bibinfo {year} {2020})}\BibitemShut {NoStop}%
\bibitem [{\citenamefont {Sancho}\ \emph {et~al.}(1985)\citenamefont {Sancho},
  \citenamefont {Sancho}, \citenamefont {Sancho},\ and\ \citenamefont
  {Rubio}}]{Sancho1985}%
  \BibitemOpen
  \bibfield  {author} {\bibinfo {author} {\bibfnamefont {M.~P.~L.}\
  \bibnamefont {Sancho}}, \bibinfo {author} {\bibfnamefont {J.~M.~L.}\
  \bibnamefont {Sancho}}, \bibinfo {author} {\bibfnamefont {J.~M.~L.}\
  \bibnamefont {Sancho}}, \ and\ \bibinfo {author} {\bibfnamefont
  {J.}~\bibnamefont {Rubio}},\ }\href {\doibase 10.1088/0305-4608/15/4/009}
  {\bibfield  {journal} {\bibinfo  {journal} {J. Phys. F: Met. Phys.}\ }\textbf
  {\bibinfo {volume} {15}},\ \bibinfo {pages} {851} (\bibinfo {year}
  {1985})}\BibitemShut {NoStop}%
\bibitem [{\citenamefont {Wu}\ \emph {et~al.}(2018)\citenamefont {Wu},
  \citenamefont {Zhang}, \citenamefont {Song}, \citenamefont {Troyer},\ and\
  \citenamefont {Soluyanov}}]{Wu2018}%
  \BibitemOpen
  \bibfield  {author} {\bibinfo {author} {\bibfnamefont {Q.}~\bibnamefont
  {Wu}}, \bibinfo {author} {\bibfnamefont {S.}~\bibnamefont {Zhang}}, \bibinfo
  {author} {\bibfnamefont {H.-F.}\ \bibnamefont {Song}}, \bibinfo {author}
  {\bibfnamefont {M.}~\bibnamefont {Troyer}}, \ and\ \bibinfo {author}
  {\bibfnamefont {A.~A.}\ \bibnamefont {Soluyanov}},\ }\href {\doibase
  10.1016/j.cpc.2017.09.033} {\bibfield  {journal} {\bibinfo  {journal}
  {Comput. Phys. Commun.}\ }\textbf {\bibinfo {volume} {224}},\ \bibinfo
  {pages} {405} (\bibinfo {year} {2018})}\BibitemShut {NoStop}%
\bibitem [{\citenamefont {Zhou}\ \emph {et~al.}(2018)\citenamefont {Zhou},
  \citenamefont {Lin}, \citenamefont {Huang}, \citenamefont {Zhou},
  \citenamefont {Chen}, \citenamefont {Xia}, \citenamefont {Wang},
  \citenamefont {Xie}, \citenamefont {Yu}, \citenamefont {Lei}, \citenamefont
  {Wu}, \citenamefont {Liu}, \citenamefont {Fu}, \citenamefont {Zeng},
  \citenamefont {Hsu}, \citenamefont {Yang}, \citenamefont {Lu}, \citenamefont
  {Yu}, \citenamefont {Shen}, \citenamefont {Lin}, \citenamefont {Yakobson},
  \citenamefont {Liu}, \citenamefont {Suenaga}, \citenamefont {Liu},\ and\
  \citenamefont {Liu}}]{zhou2018library}%
  \BibitemOpen
  \bibfield  {author} {\bibinfo {author} {\bibfnamefont {J.}~\bibnamefont
  {Zhou}}, \bibinfo {author} {\bibfnamefont {J.}~\bibnamefont {Lin}}, \bibinfo
  {author} {\bibfnamefont {X.}~\bibnamefont {Huang}}, \bibinfo {author}
  {\bibfnamefont {Y.}~\bibnamefont {Zhou}}, \bibinfo {author} {\bibfnamefont
  {Y.}~\bibnamefont {Chen}}, \bibinfo {author} {\bibfnamefont {J.}~\bibnamefont
  {Xia}}, \bibinfo {author} {\bibfnamefont {H.}~\bibnamefont {Wang}}, \bibinfo
  {author} {\bibfnamefont {Y.}~\bibnamefont {Xie}}, \bibinfo {author}
  {\bibfnamefont {H.}~\bibnamefont {Yu}}, \bibinfo {author} {\bibfnamefont
  {J.}~\bibnamefont {Lei}}, \bibinfo {author} {\bibfnamefont {D.}~\bibnamefont
  {Wu}}, \bibinfo {author} {\bibfnamefont {F.}~\bibnamefont {Liu}}, \bibinfo
  {author} {\bibfnamefont {Q.}~\bibnamefont {Fu}}, \bibinfo {author}
  {\bibfnamefont {Q.}~\bibnamefont {Zeng}}, \bibinfo {author} {\bibfnamefont
  {C.-H.}\ \bibnamefont {Hsu}}, \bibinfo {author} {\bibfnamefont
  {C.}~\bibnamefont {Yang}}, \bibinfo {author} {\bibfnamefont {L.}~\bibnamefont
  {Lu}}, \bibinfo {author} {\bibfnamefont {T.}~\bibnamefont {Yu}}, \bibinfo
  {author} {\bibfnamefont {Z.}~\bibnamefont {Shen}}, \bibinfo {author}
  {\bibfnamefont {H.}~\bibnamefont {Lin}}, \bibinfo {author} {\bibfnamefont
  {B.~I.}\ \bibnamefont {Yakobson}}, \bibinfo {author} {\bibfnamefont
  {Q.}~\bibnamefont {Liu}}, \bibinfo {author} {\bibfnamefont {K.}~\bibnamefont
  {Suenaga}}, \bibinfo {author} {\bibfnamefont {G.}~\bibnamefont {Liu}}, \ and\
  \bibinfo {author} {\bibfnamefont {Z.}~\bibnamefont {Liu}},\ }\href {\doibase
  10.1038/s41586-018-0008-3} {\bibfield  {journal} {\bibinfo  {journal}
  {Nature}\ }\textbf {\bibinfo {volume} {556}},\ \bibinfo {pages} {355}
  (\bibinfo {year} {2018})}\BibitemShut {NoStop}%
\bibitem [{\citenamefont {Zhu}\ \emph {et~al.}(2020)\citenamefont {Zhu},
  \citenamefont {Wang}, \citenamefont {Jing}, \citenamefont {Heine},\ and\
  \citenamefont {Li}}]{Zhu2020PdBi}%
  \BibitemOpen
  \bibfield  {author} {\bibinfo {author} {\bibfnamefont {X.}~\bibnamefont
  {Zhu}}, \bibinfo {author} {\bibfnamefont {Y.}~\bibnamefont {Wang}}, \bibinfo
  {author} {\bibfnamefont {Y.}~\bibnamefont {Jing}}, \bibinfo {author}
  {\bibfnamefont {T.}~\bibnamefont {Heine}}, \ and\ \bibinfo {author}
  {\bibfnamefont {Y.}~\bibnamefont {Li}},\ }\href {\doibase
  10.1016/j.mtadv.2020.100091} {\bibfield  {journal} {\bibinfo  {journal}
  {Mater. Today Adv.}\ }\textbf {\bibinfo {volume} {8}},\ \bibinfo {pages}
  {100091} (\bibinfo {year} {2020})}\BibitemShut {NoStop}%
\bibitem [{\citenamefont {Heyd}\ \emph {et~al.}(2003)\citenamefont {Heyd},
  \citenamefont {Scuseria},\ and\ \citenamefont {Ernzerhof}}]{Heyd2003}%
  \BibitemOpen
  \bibfield  {author} {\bibinfo {author} {\bibfnamefont {J.}~\bibnamefont
  {Heyd}}, \bibinfo {author} {\bibfnamefont {G.~E.}\ \bibnamefont {Scuseria}},
  \ and\ \bibinfo {author} {\bibfnamefont {M.}~\bibnamefont {Ernzerhof}},\
  }\href {\doibase 10.1063/1.1564060} {\bibfield  {journal} {\bibinfo
  {journal} {J. Chem. Phys.}\ }\textbf {\bibinfo {volume} {118}},\ \bibinfo
  {pages} {8207} (\bibinfo {year} {2003})}\BibitemShut {NoStop}%
\bibitem [{\citenamefont {Zhao}\ \emph {et~al.}(2018)\citenamefont {Zhao},
  \citenamefont {Zeng}, \citenamefont {Lian}, \citenamefont {Dai},
  \citenamefont {Meng},\ and\ \citenamefont {Ni}}]{Zhao2018}%
  \BibitemOpen
  \bibfield  {author} {\bibinfo {author} {\bibfnamefont {Y.}~\bibnamefont
  {Zhao}}, \bibinfo {author} {\bibfnamefont {S.}~\bibnamefont {Zeng}}, \bibinfo
  {author} {\bibfnamefont {C.}~\bibnamefont {Lian}}, \bibinfo {author}
  {\bibfnamefont {Z.}~\bibnamefont {Dai}}, \bibinfo {author} {\bibfnamefont
  {S.}~\bibnamefont {Meng}}, \ and\ \bibinfo {author} {\bibfnamefont
  {J.}~\bibnamefont {Ni}},\ }\href {\doibase 10.1103/PhysRevB.98.134514}
  {\bibfield  {journal} {\bibinfo  {journal} {Phys. Rev. B}\ }\textbf {\bibinfo
  {volume} {98}},\ \bibinfo {pages} {134514} (\bibinfo {year}
  {2018})}\BibitemShut {NoStop}%
\bibitem [{\citenamefont {Yan}\ \emph {et~al.}(2019)\citenamefont {Yan},
  \citenamefont {Liu}, \citenamefont {Bo}, \citenamefont {Zhang}, \citenamefont
  {Tang}, \citenamefont {Xiao},\ and\ \citenamefont {Wang}}]{Yan2019}%
  \BibitemOpen
  \bibfield  {author} {\bibinfo {author} {\bibfnamefont {L.}~\bibnamefont
  {Yan}}, \bibinfo {author} {\bibfnamefont {P.-F.}\ \bibnamefont {Liu}},
  \bibinfo {author} {\bibfnamefont {T.}~\bibnamefont {Bo}}, \bibinfo {author}
  {\bibfnamefont {J.}~\bibnamefont {Zhang}}, \bibinfo {author} {\bibfnamefont
  {M.-H.}\ \bibnamefont {Tang}}, \bibinfo {author} {\bibfnamefont {Y.-G.}\
  \bibnamefont {Xiao}}, \ and\ \bibinfo {author} {\bibfnamefont {B.-T.}\
  \bibnamefont {Wang}},\ }\href {\doibase 10.1039/c9tc03740c} {\bibfield
  {journal} {\bibinfo  {journal} {J. Mater. Chem. C}\ }\textbf {\bibinfo
  {volume} {7}},\ \bibinfo {pages} {10926} (\bibinfo {year}
  {2019})}\BibitemShut {NoStop}%
\bibitem [{\citenamefont {Zhang}\ and\ \citenamefont {Dong}(2017)}]{Zhang2017}%
  \BibitemOpen
  \bibfield  {author} {\bibinfo {author} {\bibfnamefont {J.-J.}\ \bibnamefont
  {Zhang}}\ and\ \bibinfo {author} {\bibfnamefont {S.}~\bibnamefont {Dong}},\
  }\href {\doibase 10.1063/1.4974085} {\bibfield  {journal} {\bibinfo
  {journal} {J. Chem. Phys.}\ }\textbf {\bibinfo {volume} {146}},\ \bibinfo
  {pages} {034705} (\bibinfo {year} {2017})}\BibitemShut {NoStop}%
\bibitem [{\citenamefont {Calandra}\ and\ \citenamefont
  {Mauri}(2011)}]{Calandra2011}%
  \BibitemOpen
  \bibfield  {author} {\bibinfo {author} {\bibfnamefont {M.}~\bibnamefont
  {Calandra}}\ and\ \bibinfo {author} {\bibfnamefont {F.}~\bibnamefont
  {Mauri}},\ }\href {\doibase 10.1103/PhysRevLett.106.196406} {\bibfield
  {journal} {\bibinfo  {journal} {Phys. Rev. Lett.}\ }\textbf {\bibinfo
  {volume} {106}},\ \bibinfo {pages} {196406} (\bibinfo {year}
  {2011})}\BibitemShut {NoStop}%
\bibitem [{\citenamefont {Zhuang}\ \emph {et~al.}(2017)\citenamefont {Zhuang},
  \citenamefont {Johannes}, \citenamefont {Singh},\ and\ \citenamefont
  {Hennig}}]{Zhuang2017}%
  \BibitemOpen
  \bibfield  {author} {\bibinfo {author} {\bibfnamefont {H.~L.}\ \bibnamefont
  {Zhuang}}, \bibinfo {author} {\bibfnamefont {M.~D.}\ \bibnamefont
  {Johannes}}, \bibinfo {author} {\bibfnamefont {A.~K.}\ \bibnamefont {Singh}},
  \ and\ \bibinfo {author} {\bibfnamefont {R.~G.}\ \bibnamefont {Hennig}},\
  }\href {\doibase 10.1103/PhysRevB.96.165305} {\bibfield  {journal} {\bibinfo
  {journal} {Phy. Rev. B}\ }\textbf {\bibinfo {volume} {96}},\ \bibinfo {pages}
  {165305} (\bibinfo {year} {2017})}\BibitemShut {NoStop}%
\bibitem [{\citenamefont {Lv}\ \emph {et~al.}(2017)\citenamefont {Lv},
  \citenamefont {Wang}, \citenamefont {Zhang}, \citenamefont {Ding},
  \citenamefont {Li}, \citenamefont {Wang}, \citenamefont {He}, \citenamefont
  {Song}, \citenamefont {Ma},\ and\ \citenamefont {Xue}}]{LV2017852}%
  \BibitemOpen
  \bibfield  {author} {\bibinfo {author} {\bibfnamefont {Y.-F.}\ \bibnamefont
  {Lv}}, \bibinfo {author} {\bibfnamefont {W.-L.}\ \bibnamefont {Wang}},
  \bibinfo {author} {\bibfnamefont {Y.-M.}\ \bibnamefont {Zhang}}, \bibinfo
  {author} {\bibfnamefont {H.}~\bibnamefont {Ding}}, \bibinfo {author}
  {\bibfnamefont {W.}~\bibnamefont {Li}}, \bibinfo {author} {\bibfnamefont
  {L.}~\bibnamefont {Wang}}, \bibinfo {author} {\bibfnamefont {K.}~\bibnamefont
  {He}}, \bibinfo {author} {\bibfnamefont {C.-L.}\ \bibnamefont {Song}},
  \bibinfo {author} {\bibfnamefont {X.-C.}\ \bibnamefont {Ma}}, \ and\ \bibinfo
  {author} {\bibfnamefont {Q.-K.}\ \bibnamefont {Xue}},\ }\href {\doibase
  https://doi.org/10.1016/j.scib.2017.05.008} {\bibfield  {journal} {\bibinfo
  {journal} {Sci. Bull.}\ }\textbf {\bibinfo {volume} {62}},\ \bibinfo {pages}
  {852 } (\bibinfo {year} {2017})}\BibitemShut {NoStop}%
\bibitem [{\citenamefont {Schoop}\ \emph {et~al.}(2015)\citenamefont {Schoop},
  \citenamefont {Xie}, \citenamefont {Chen}, \citenamefont {Gibson},
  \citenamefont {Lapidus}, \citenamefont {Kimchi}, \citenamefont
  {Hirschberger}, \citenamefont {Haldolaarachchige}, \citenamefont {Ali},
  \citenamefont {Belvin}, \citenamefont {Liang}, \citenamefont {Neaton},
  \citenamefont {Ong}, \citenamefont {Vishwanath},\ and\ \citenamefont
  {Cava}}]{PhysRevB.91.214517}%
  \BibitemOpen
  \bibfield  {author} {\bibinfo {author} {\bibfnamefont {L.~M.}\ \bibnamefont
  {Schoop}}, \bibinfo {author} {\bibfnamefont {L.~S.}\ \bibnamefont {Xie}},
  \bibinfo {author} {\bibfnamefont {R.}~\bibnamefont {Chen}}, \bibinfo {author}
  {\bibfnamefont {Q.~D.}\ \bibnamefont {Gibson}}, \bibinfo {author}
  {\bibfnamefont {S.~H.}\ \bibnamefont {Lapidus}}, \bibinfo {author}
  {\bibfnamefont {I.}~\bibnamefont {Kimchi}}, \bibinfo {author} {\bibfnamefont
  {M.}~\bibnamefont {Hirschberger}}, \bibinfo {author} {\bibfnamefont
  {N.}~\bibnamefont {Haldolaarachchige}}, \bibinfo {author} {\bibfnamefont
  {M.~N.}\ \bibnamefont {Ali}}, \bibinfo {author} {\bibfnamefont {C.~A.}\
  \bibnamefont {Belvin}}, \bibinfo {author} {\bibfnamefont {T.}~\bibnamefont
  {Liang}}, \bibinfo {author} {\bibfnamefont {J.~B.}\ \bibnamefont {Neaton}},
  \bibinfo {author} {\bibfnamefont {N.~P.}\ \bibnamefont {Ong}}, \bibinfo
  {author} {\bibfnamefont {A.}~\bibnamefont {Vishwanath}}, \ and\ \bibinfo
  {author} {\bibfnamefont {R.~J.}\ \bibnamefont {Cava}},\ }\href {\doibase
  10.1103/PhysRevB.91.214517} {\bibfield  {journal} {\bibinfo  {journal} {Phys.
  Rev. B}\ }\textbf {\bibinfo {volume} {91}},\ \bibinfo {pages} {214517}
  (\bibinfo {year} {2015})}\BibitemShut {NoStop}%
\bibitem [{\citenamefont {Xing}\ \emph {et~al.}(2016)\citenamefont {Xing},
  \citenamefont {Wang}, \citenamefont {Li}, \citenamefont {Zhang},
  \citenamefont {Liu}, \citenamefont {Zhang}, \citenamefont {Luo},
  \citenamefont {Wang}, \citenamefont {Wang}, \citenamefont {Ling},
  \citenamefont {Tian}, \citenamefont {Jia}, \citenamefont {Feng},
  \citenamefont {Liu}, \citenamefont {Wei},\ and\ \citenamefont
  {Wang}}]{Xing2016}%
  \BibitemOpen
  \bibfield  {author} {\bibinfo {author} {\bibfnamefont {Y.}~\bibnamefont
  {Xing}}, \bibinfo {author} {\bibfnamefont {H.}~\bibnamefont {Wang}}, \bibinfo
  {author} {\bibfnamefont {C.-K.}\ \bibnamefont {Li}}, \bibinfo {author}
  {\bibfnamefont {X.}~\bibnamefont {Zhang}}, \bibinfo {author} {\bibfnamefont
  {J.}~\bibnamefont {Liu}}, \bibinfo {author} {\bibfnamefont {Y.}~\bibnamefont
  {Zhang}}, \bibinfo {author} {\bibfnamefont {J.}~\bibnamefont {Luo}}, \bibinfo
  {author} {\bibfnamefont {Z.}~\bibnamefont {Wang}}, \bibinfo {author}
  {\bibfnamefont {Y.}~\bibnamefont {Wang}}, \bibinfo {author} {\bibfnamefont
  {L.}~\bibnamefont {Ling}}, \bibinfo {author} {\bibfnamefont {M.}~\bibnamefont
  {Tian}}, \bibinfo {author} {\bibfnamefont {S.}~\bibnamefont {Jia}}, \bibinfo
  {author} {\bibfnamefont {J.}~\bibnamefont {Feng}}, \bibinfo {author}
  {\bibfnamefont {X.-J.}\ \bibnamefont {Liu}}, \bibinfo {author} {\bibfnamefont
  {J.}~\bibnamefont {Wei}}, \ and\ \bibinfo {author} {\bibfnamefont
  {J.}~\bibnamefont {Wang}},\ }\href {\doibase 10.1038/npjquantmats.2016.5}
  {\bibfield  {journal} {\bibinfo  {journal} {npj Quantum Mater.}\ }\textbf
  {\bibinfo {volume} {1}},\ \bibinfo {pages} {1} (\bibinfo {year}
  {2016})}\BibitemShut {NoStop}%
\bibitem [{\citenamefont {Fu}\ and\ \citenamefont
  {Kane}(2007)}]{PhysRevB.76.045302}%
  \BibitemOpen
  \bibfield  {author} {\bibinfo {author} {\bibfnamefont {L.}~\bibnamefont
  {Fu}}\ and\ \bibinfo {author} {\bibfnamefont {C.~L.}\ \bibnamefont {Kane}},\
  }\href {\doibase 10.1103/PhysRevB.76.045302} {\bibfield  {journal} {\bibinfo
  {journal} {Phys. Rev. B}\ }\textbf {\bibinfo {volume} {76}},\ \bibinfo
  {pages} {045302} (\bibinfo {year} {2007})}\BibitemShut {NoStop}%
\bibitem [{\citenamefont {Fu}\ \emph {et~al.}(2007)\citenamefont {Fu},
  \citenamefont {Kane},\ and\ \citenamefont {Mele}}]{PhysRevLett.98.106803}%
  \BibitemOpen
  \bibfield  {author} {\bibinfo {author} {\bibfnamefont {L.}~\bibnamefont
  {Fu}}, \bibinfo {author} {\bibfnamefont {C.~L.}\ \bibnamefont {Kane}}, \ and\
  \bibinfo {author} {\bibfnamefont {E.~J.}\ \bibnamefont {Mele}},\ }\href
  {\doibase 10.1103/PhysRevLett.98.106803} {\bibfield  {journal} {\bibinfo
  {journal} {Phys. Rev. Lett.}\ }\textbf {\bibinfo {volume} {98}},\ \bibinfo
  {pages} {106803} (\bibinfo {year} {2007})}\BibitemShut {NoStop}%
\bibitem [{\citenamefont {Qi}\ \emph {et~al.}(2010)\citenamefont {Qi},
  \citenamefont {Hughes},\ and\ \citenamefont {Zhang}}]{PhysRevB.81.134508}%
  \BibitemOpen
  \bibfield  {author} {\bibinfo {author} {\bibfnamefont {X.-L.}\ \bibnamefont
  {Qi}}, \bibinfo {author} {\bibfnamefont {T.~L.}\ \bibnamefont {Hughes}}, \
  and\ \bibinfo {author} {\bibfnamefont {S.-C.}\ \bibnamefont {Zhang}},\ }\href
  {\doibase 10.1103/PhysRevB.81.134508} {\bibfield  {journal} {\bibinfo
  {journal} {Phys. Rev. B}\ }\textbf {\bibinfo {volume} {81}},\ \bibinfo
  {pages} {134508} (\bibinfo {year} {2010})}\BibitemShut {NoStop}%
\bibitem [{\citenamefont {Jin}\ \emph {et~al.}(2019)\citenamefont {Jin},
  \citenamefont {Huang}, \citenamefont {Mei}, \citenamefont {Liu},
  \citenamefont {Lim},\ and\ \citenamefont {Liu}}]{Jin2019MgB}%
  \BibitemOpen
  \bibfield  {author} {\bibinfo {author} {\bibfnamefont {K.-H.}\ \bibnamefont
  {Jin}}, \bibinfo {author} {\bibfnamefont {H.}~\bibnamefont {Huang}}, \bibinfo
  {author} {\bibfnamefont {J.-W.}\ \bibnamefont {Mei}}, \bibinfo {author}
  {\bibfnamefont {Z.}~\bibnamefont {Liu}}, \bibinfo {author} {\bibfnamefont
  {L.-K.}\ \bibnamefont {Lim}}, \ and\ \bibinfo {author} {\bibfnamefont
  {F.}~\bibnamefont {Liu}},\ }\href {\doibase 10.1038/s41524-019-0191-2}
  {\bibfield  {journal} {\bibinfo  {journal} {npj Comput. Mater.}\ }\textbf
  {\bibinfo {volume} {5}},\ \bibinfo {pages} {1} (\bibinfo {year}
  {2019})}\BibitemShut {NoStop}%
\bibitem [{\citenamefont {Sato}(2003)}]{Sato2003}%
  \BibitemOpen
  \bibfield  {author} {\bibinfo {author} {\bibfnamefont {M.}~\bibnamefont
  {Sato}},\ }\href {\doibase 10.1016/j.physletb.2003.09.047} {\bibfield
  {journal} {\bibinfo  {journal} {Phys. Lett. B}\ }\textbf {\bibinfo {volume}
  {575}},\ \bibinfo {pages} {126} (\bibinfo {year} {2003})}\BibitemShut
  {NoStop}%
\bibitem [{\citenamefont {Fu}\ and\ \citenamefont
  {Kane}(2008)}]{PhysRevLett.100.096407}%
  \BibitemOpen
  \bibfield  {author} {\bibinfo {author} {\bibfnamefont {L.}~\bibnamefont
  {Fu}}\ and\ \bibinfo {author} {\bibfnamefont {C.~L.}\ \bibnamefont {Kane}},\
  }\href {\doibase 10.1103/PhysRevLett.100.096407} {\bibfield  {journal}
  {\bibinfo  {journal} {Phys. Rev. Lett.}\ }\textbf {\bibinfo {volume} {100}},\
  \bibinfo {pages} {096407} (\bibinfo {year} {2008})}\BibitemShut {NoStop}%
\bibitem [{\citenamefont {Tu}\ \emph {et~al.}(2020)\citenamefont {Tu},
  \citenamefont {Liu}, \citenamefont {Yin}, \citenamefont {Zhang},\ and\
  \citenamefont {Wang}}]{2008.06260}%
  \BibitemOpen
  \bibfield  {author} {\bibinfo {author} {\bibfnamefont {X.-H.}\ \bibnamefont
  {Tu}}, \bibinfo {author} {\bibfnamefont {P.-F.}\ \bibnamefont {Liu}},
  \bibinfo {author} {\bibfnamefont {W.}~\bibnamefont {Yin}}, \bibinfo {author}
  {\bibfnamefont {J.-R.}\ \bibnamefont {Zhang}}, \ and\ \bibinfo {author}
  {\bibfnamefont {B.-T.}\ \bibnamefont {Wang}},\ }\href@noop {} {\enquote
  {\bibinfo {title} {Schemes for realizing topological superconductivity in
  monolayer $\beta$-{B}i$_2${P}d: A tight-binding model study},}\ } (\bibinfo
  {year} {2020}),\ \Eprint {http://arxiv.org/abs/arXiv:2008.06260}
  {arXiv:2008.06260} \BibitemShut {NoStop}%
\bibitem [{\citenamefont {Qiao}\ \emph {et~al.}(2014)\citenamefont {Qiao},
  \citenamefont {Ren}, \citenamefont {Chen}, \citenamefont {Bellaiche},
  \citenamefont {Zhang}, \citenamefont {MacDonald},\ and\ \citenamefont
  {Niu}}]{PhysRevLett.112.116404}%
  \BibitemOpen
  \bibfield  {author} {\bibinfo {author} {\bibfnamefont {Z.}~\bibnamefont
  {Qiao}}, \bibinfo {author} {\bibfnamefont {W.}~\bibnamefont {Ren}}, \bibinfo
  {author} {\bibfnamefont {H.}~\bibnamefont {Chen}}, \bibinfo {author}
  {\bibfnamefont {L.}~\bibnamefont {Bellaiche}}, \bibinfo {author}
  {\bibfnamefont {Z.}~\bibnamefont {Zhang}}, \bibinfo {author} {\bibfnamefont
  {A.~H.}\ \bibnamefont {MacDonald}}, \ and\ \bibinfo {author} {\bibfnamefont
  {Q.}~\bibnamefont {Niu}},\ }\href {\doibase 10.1103/PhysRevLett.112.116404}
  {\bibfield  {journal} {\bibinfo  {journal} {Phys. Rev. Lett.}\ }\textbf
  {\bibinfo {volume} {112}},\ \bibinfo {pages} {116404} (\bibinfo {year}
  {2014})}\BibitemShut {NoStop}%
\end{thebibliography}
\end{document}